\documentclass[lettersize,journal]{IEEEtran}
\usepackage{amsmath,amsfonts}
\usepackage{algorithmic}
\usepackage{algorithm}
\usepackage{array}
\usepackage[caption=false,font=normalsize,labelfont=sf,textfont=sf]{subfig}
\usepackage{textcomp}
\usepackage{stfloats}
\usepackage{url}
\usepackage{verbatim}
\usepackage{graphicx}
\usepackage{cite}
\usepackage{comment}
\usepackage{amssymb}
\usepackage{amsmath}
\usepackage{graphicx}
\usepackage{amsthm}
\usepackage{multicol}
\usepackage{balance}
\usepackage{lineno}
\theoremstyle{remark}
\newtheorem{lemma}{Lemma}
\newtheorem{remark}{Remark}
\newtheorem{condition}{Condition}
\newtheorem{prop}{Proposition}
\usepackage{enumitem}
\usepackage{xcolor} 
\usepackage{soul}
\sethlcolor{white}

\begin{document}

\title{Dynamic Input Mapping Inversion to Eliminate Algebraic Loops in Hydraulic Actuator Control}

\author{Alessio Dallabona, Patrik Schermann, Mogens Blanke, Dimitrios Papageorgiou
\thanks{The authors are with DTU, 2800 Lyngby, Denmark. Corresponding author: Dallabona A. (aldall@dtu.dk).}
\thanks{© 2025 IEEE. This is the author version of an article published in \textit{IEEE Transactions on Control Systems Technology}. The final version is available at DOI: 10.1109/TCST.2025.3641638.}

}



\maketitle

\begin{abstract}
The application of nonlinear control schemes to electro-hydraulic actuators often requires several alterations in the design of the controllers during their implementation. This is to overcome challenges that frequently arise in such control algorithms owing to model nonlinearities. Moreover, advanced control solutions for this type of systems often introduce input algebraic loops that pose significant design and tuning difficulties. Conventional methods to avoid such loops introduce chatter, which considerably degrade tracking performance and has oil degradation and wear as side effects. This study presents a nonlinear control architecture for hydraulic actuators that comprises low-complexity modules  that facilitate robust high performance in tracking and avoids the drawbacks of chatter. The salient feature is a dynamic input-mapping inversion module that avoids algebraic loops in the control input and is followed by dedicated position control. The stability of the closed-loop system is analyzed using arguments from Lyapunov theory for cascaded non-autonomous nonlinear systems. The effectiveness of the proposed solution is evaluated on a high-fidelity simulator of a wind turbine pitch system, and validated on a full-scale laboratory setup that includes a hydraulic pitch system and blade bearing. Appropriate quantitative metrics are used to evaluate the closed-loop system performance in comparison to a state-of-the-art nonlinear design.
\end{abstract}

\begin{IEEEkeywords}
electro-hydraulic actuator, nonlinear control, parameter estimation, mapping inversion
\end{IEEEkeywords}

\section{Introduction}
\IEEEPARstart{E}{lectro-hydrualic} actuators (EHA) are commonly used in industrial applications because of their high power density and resilience to harsh environment. Typical applications include aviation, manufacturing, marine/offshore engineering, robotics and energy systems \cite{fossenMarineControlSystems2002,walgernReliabilityElectricalHydraulic2023,dallabonaFaultDiagnosisPrognosis2025}.

Although Proportional-Integral-Differential (PID) controllers constitute the industrial standard in these application domains, they often fall short in maintaining high performance in the presence of strong nonlinearities and uncertainties that characterize hydraulic actuator systems. This has motivated the exploration of nonlinear control alternatives that can improve performance and enhance closed-loop robustness over the entire envelope of operation. The complexity, however, of system dynamics often hinders effective application of nonlinear model-based solutions because several limiting assumptions and modeling simplifications need be imposed to apply these methods.

A common heavy simplification concerns the valve dynamics, where two main issues can be identified. A first limitation is encountered when a system model cannot be written in the affine form $\dot{x} = f(x) + g(x)u$, as the vector field depends non-linearly on the input. A common approach to address this issue is to use the input value at the previous time instant, under the assumption that the input $u$ is slowly varying. This removes the nonlinearity and avoids the introduction of an algebraic loop, at the cost of increased chatter in the control signal \cite{ayalewCascadeTuningNonlinear2006}. The second design challenge arises with inclusion of actuator dynamics, which increases the relative degree of the system, thus making the design much more complicated, as will be discussed later. Neglecting the valve dynamics during the design stage can simplify implementation but may potentially result in a quite demanding control action
going beyond obtainable valve response. This may introduce additional chatter and potential wind up effects.

Furthermore, micro-movements in EHAs  are responsible for local heating of hydraulic oil, which degrades oil quality. This leads to increased friction, and eventually to wear and potential damage of the valve \cite{dallabonaFrictionEstimationCondition2024a}. It is therefore essential to establish control designs that combine the performance and robustness features of nonlinear solutions, but avoid excessive actuation demand.

Several studies in existing literature have reported solutions that adopt nonlinear strategies for controlling hydraulic systems. A cascaded feedback linearization controller with a sliding mode observer was designed in \cite{kochObserverbasedSlidingMode2016}. Sliding mode was explored in a standard fashion in \cite{thomasPracticalProfileTracking2022}, \cite{guanAdaptiveSlidingMode2008a}, for high order sliding modes in \cite{estradaHydraulicActuatorControl2025} and in \cite{estradaSupertwistingBasedSliding2024} to eliminate the velocity state in the control action. The authors in \protect\cite{yamamotoForceControllerValvemanipulated2025} and \protect\cite{yamamotoPositionControllerHydraulic2025} combined sliding mode with a quasi-static EHA model using a set-valued formalism to capture model nonlinearities. A fixed-time Event-Triggered Control strategy has been tested in \protect\cite{wangDisturbanceObserverBasedFixedTime2025}. Adaptive solutions have been applied by means of backstepping, in e.g.  \cite{chengguanNonlinearAdaptiveRobust2008}, \cite{chouxCascadeControllerIncluding2012}, \cite{bakhshandeProportionalIntegralObserverBasedBacksteppingApproach2018}, \cite{manganasBacksteppingBasedController2023}, and neural networks in \protect\cite{guoNeuralAdaptiveBackstepping2019}, \protect\cite{yangNeuralAdaptiveDynamic2023}, \cite{phanAdaptiveNeuralObserverbased2025}.
The authors in \cite{chouxAdaptiveBacksteppingControl2010a} used a coordinate transformation and tuning functions to obtain a backstepping controller that included the valve dynamics. The high relative degree led to a complex solution that would also be affected by valve degradation over time.

Except for said article, all other published solutions adopted the strategy presented in \cite{ayalewCascadeTuningNonlinear2006} to cancel the algebraic loop, and did not account for the input valve dynamics in the design of the controller.

This paper uses results from online estimation of nonlinearly parametrized perturbations \cite{gripParameterEstimationCompensation2010b,papageorgiouRobustBacklashEstimation2019} to obtain a dynamical mapping inversion (MI) algorithm and derive the control input from a virtual input designed via classical nonlinear methods\hl{, without relying on the simplifying assumptions commonly adopted in the literature to implement nonlinear designs}. The paper shows how a modular and straightforward design is achieved, and how the MI can reconstruct the control signal without introducing algebraic loops. Moreover, it is shown how the suggested approach facilitates a systematic commissioning feature, namely, how it can accommodate valve specification parameters such that only frequencies that can actually be delivered by the valve are allowed in the control input. In this way, the performance can be improved and chatter in the input can be significantly reduced.

The contributions of this article are:
\begin{enumerate}
    \item A dynamical input mapping inversion algorithm in a modular architecture for control of hydraulic actuators that avoids algebraic loops and significantly reduces EHA micro-movements and associated wear.
    \item A systematic method to tune the dynamical mapping inversion component according to control valve dynamics.
    \item Rigorous analysis of stability/boundedness  and validation on a wind-turbine EHA system, using both a high-fidelity simulator and a full-scale physical test rig.
\end{enumerate}

The remainder of the paper is organized as follows. Sec.~\ref{sec: prob_def} presents the hydraulic actuator system and defines the problem at hand. Sec.~\ref{sec:standard_design} designs a well-established baseline controller following commonly accepted standards, and  Sec.~\ref{sec:virt_inp} proposes a new algorithm and presents features and benefits in contrast to the standard existing approaches. Stability of the closed-loop system is the subject of  Sec.~\ref{sec:stab}. Sec.~\ref{sec:case_study} presents a case study of a wind turbine pitch control, and results from validation on a full-scale electro-hydraulic test rig are shown in Sec.~\ref{sec:sim_res}. Finally, conclusions and future directions are offered in Sec.~\ref{sec:conclusions}.

\section{Problem definition}\label{sec: prob_def}

\begin{figure}[tp]
    \centering
\includegraphics[width=\columnwidth]{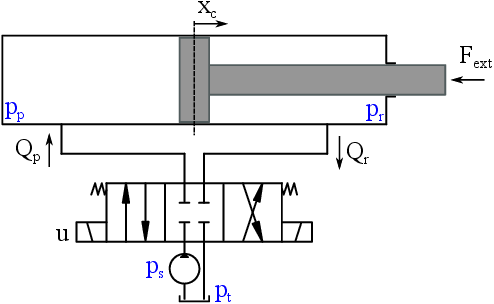}
    \caption{Generic electro-hydraulic actuator system: above, the linear cylinder with piston and rod; below, the flow control valve, oil pump and drain.}
    \label{fig:hyd_sys}
\end{figure}

An electro-hydraulic actuator is composed of an hydraulic cylinder driven by a directional control valve (DCV), as shown in Fig.~\ref{fig:hyd_sys}. The pressures $p_p$ and $p_r$ in the piston-side and rod-side chambers of the cylinder are varied according to the valve's spool position $x_v$. This allows for control of both position and  velocity ($x_c$ and $v_c$) of the  the cylinder's piston. The model for a generic electro-hydraulic actuator is given by the dynamic model \ref{eq:gen_dyn}, see e.g. \cite{komstaIntegralSlidingMode2013,dallabonaFaultDiagnosisPrognosis2025,pedersenInvestigationLoadReduction2016}:

\begin{equation}\label{eq:gen_dyn}
\begin{aligned}
& M \dot{v}_c=A_p p_p-A_r p_r-F_{F r}-F_{e x t} \\
& \dot{x_c}=v_c \\
& \dot{p}_p=\frac{B_{e, p}(p_p)}{V_p(x_c)}\left[Q_p(x_v)-A_p v_c\right] \\
& \dot{p}_r=\frac{B_{e, r}(p_r)}{V_r(x_c)}\left[-Q_r(x_v)+A_r v_c\right] \\
& \dot{v}_v+2 \xi \omega_0 v_v+\omega_0^2 x_v=\omega_0^2 u_x\\
& \dot{x_v} = v_v,
\end{aligned}
\end{equation}

where $M$ is the piston mass, $A_p$ and $A_r$ the cylinder cross sections in the two chambers, $F_{fr}$ and $F_{ext}$ the friction force and the external load respectively, and $B_{e, i}(p_i)$ the effective bulk modulus of the oil in the $i$-th component. The terms
\begin{equation}\label{eq:volumes}
\begin{aligned}
& V_p(x_c) = V_{0, p}+A_p x_c\\
& V_r(x_v) = V_{0, r}+A_r\left(x_{c, m}-x_c\right).
\end{aligned}
\end{equation}
represent the volumes of the cylinder chambers.

The flows are calculated as
\begin{equation}\label{eq:flows}
\begin{aligned}
& Q_p(x_v) = K(x_v)\sqrt{\Delta_{pp}(x_v)}\\
& Q_r(x_v) = K(x_v)\sqrt{\Delta_{pr}(x_v)}.
\end{aligned}
\end{equation}
where the nonlinear function $K(x_v)$ determine the valve flow coefficient based on its spool position, and
\begin{equation}\label{eq:delta_p}
\begin{aligned}
& \Delta_{pp} = (p_s-p_p)\,H\left(x_v\right) + (p_p-p_t)\,H\left(-x_v\right)\\
& \Delta_{pr} = (p_r-p_t)\,H\left(x_v\right) + (p_s-p_r)\,H\left(-x_v\right),
\end{aligned}
\end{equation}
where $H(\cdot)$ stands for the Heaviside step function, and $p_s$ and $p_t$ for system and tank pressures, respectively.

The terms $\omega_0$ and $\xi$ denote the natural frequency and damping ratio of the input valve, which is modeled as a second-order system. This stems from DCVs typically being equipped with an internal closed loop. As a result, the control input $u_x$ is a normalized, thus dimensionless, position reference for the valve spool. 

The control objective in EHA systems is the tracking of a reference signal for the cylinder piston position $x_c$. By inspecting the cylinder position dynamics in \eqref{eq:gen_dyn} it is easily seen that the system corresponding to the position tracking error $e$, 
\begin{equation}\label{eq:error_def}
e = x_c - r,
\end{equation}
has relative degree $\alpha = 5$, i.e. $e$ needs to be differentiated five times in order to arrive to a controller canonical form
\begin{equation}\label{eq:gen_control}
    e^{(\alpha)}= f(u),
\end{equation}
which is suitable for control design.

For the class of EHA systems this introduces a high degree of complexity due to the strong nonlinear couplings between the various pressure terms. A common approach adopted in the literature to alleviate this complexity is the introduction of a simplifying assumption, namely neglecting the valve dynamics. This implies that the position $x_v$ inside the terms $Q_p(x_v)$ and $Q_r(x_v)$ can instantaneously assume any desired value and is, therefore, considered the effective control input for the system, i.e. $u \triangleq x_v$ and $Q_p(x_v),Q_r(x_v)$ becoming $Q_p(u),Q_r(u)$.

An additional frequent simplification made in many reported studies is considering one single constant value for the bulk modulus of the oil for both chambers, i.e., $B_{e, p}=B_{e, r}=B > 0$. The resulting model, which is most commonly used in the industry for control design, is \cite{bakhshandeRobustControlApproach2017}:

\begin{equation}\label{eq:final_model}
\begin{aligned}
& M \dot{v}_c = A_p p_p - A_r p_r - F_{Fr} - F_{ext} \\
& \dot{x}_c = v_c \\
& \dot{F}_h = A_p \dot{p}_p - A_r \dot{p}_r \\
& \quad = \frac{A_p B}{V_{p}(x_c)} \big[ Q_p(u) - A_p v_c \big]  - \frac{A_r B}{V_{r}(x_c)} \big[ A_r v_c - Q_r(u) \big],
\end{aligned}
\end{equation}
where the hydraulic force $F_h = A_pp_p-A_rp_r$ is introduced to avoid considering internal dynamics.

\section{Baseline nonlinear control design}\label{sec:standard_design}

\begin{figure}[b]
    \centering
    \includegraphics[width=0.9\columnwidth]{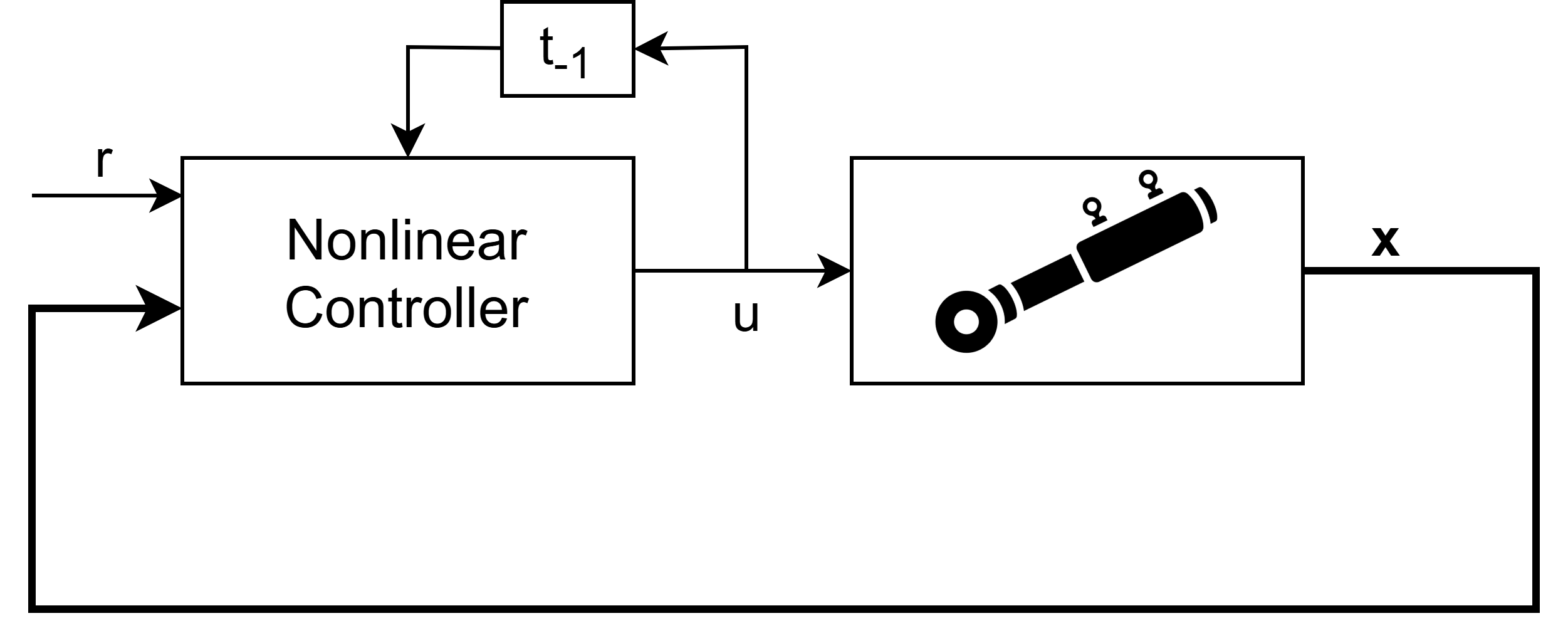}
    \caption{Block diagram of the baseline algorithm that does not consider DCV dynamics.}
    \label{fig:block_diag_base}
\end{figure}

The control architecture for the baseline solution is presented in Fig.~\ref{fig:block_diag_base}, and represent the standard configuration for nonlinear control solutions for the system under analysis.

Considering the model presented in  \eqref{eq:final_model} and the tracking error $e$ defined in \eqref{eq:error_def}, the dynamics constitutes a relative degree 3 system and reads,

\begin{equation}\label{eq:ddde}
\begin{aligned}
e^{(3)} =  &-\frac{\dot{F}_{fric} + \dot{F}_{ext}}{M} - \frac{A_p^2B}{MV_p}v_c - \frac{A_r^2B}{MV_r}v_c - r^{(3)} +\\
&\frac{A_pB}{MV_p}Q_p(u) + \frac{A_rB}{MV_r}Q_r(u) \triangleq  \, \varpi.
\end{aligned}
\end{equation}
where the \hl{term $\varpi = \varpi(t,\boldsymbol{x},u)$} is a function of time, of states $\boldsymbol{x} \triangleq \begin{bmatrix}
    v_c & x_c & F_h 
\end{bmatrix}^T$, and of the control input $u$. 

Intermediate derivatives are given by

\begin{equation}\label{eq:errors}
\begin{aligned}
    \dot{e} &= v_{c} - \dot{r} \\
    \ddot{e} &= -\frac{F_{fr} + F_{ext}}{M} + \frac{A_p p_p - A_r p_r}{M} - \ddot{r},
\end{aligned}
\end{equation}

Feedback linearization (FL) is chosen as the baseline nonlinear control design to showcase the benefits of the new developed algorithm, given its widespread adoption in adaptive control schemes. Other solutions like sliding mode control, which often rely on an equivalent control law for satisfactory performance\cite{dallabonaModelfreeSlidingMode2024}, could similarly be used.


A feedback linearization (FL) control is obtained from writing the error derivative dynamics as follows,
\begin{equation}
	\dot{\underline{\mathbf{e}}} = 
	\underbrace{\begin{bmatrix}
			0 & 1 & 0\\
			0 & 0 & 1\\
            0 & 0 & 0
	\end{bmatrix}}_{\boldsymbol{A_{FL}}} \underbrace{\begin{bmatrix}
	    e\\ \dot{e} \\ \ddot{e}
	\end{bmatrix}}_{\underline{\mathbf{e}}}  +
	\underbrace{\begin{bmatrix}
			0\\
            0\\
			1
	\end{bmatrix}}_{\boldsymbol{B_{FL}}}
	\varpi.
 \label{eq:fl_dynamics}
\end{equation}
\hl{Selecting $\varpi$ as} 
\begin{equation}\label{eq:vd}
    \varpi = -K_{FL}\underline{\mathbf{e}},
\end{equation}
with $K_{FL}$ obtained by a standard linear design (eigenvalue assignment, LQR, or other), renders the closed-loop system globally exponentially stable (GES) with the rate of decay being defined by the gain $K_{FL}$. The related control input $u$ is isolated from \eqref{eq:vd} taking also \eqref{eq:ddde}, and \eqref{eq:errors} into account.

From the latter it is noticed that the error dynamics depends on the terms $F_{ext}$ and $F_{fr}$, and their derivatives. Such terms are in general unknown and need be estimated to later compute the control input $u$. Although dedicated offline tests could  identify the friction coefficients as they are during commissioning, the friction will change over time and it is both convenient and necessary to assume friction to be unknown. This allows for lumping friction together with external load as a single term $F_d = F_{ext} + F_{fr}$, which can  be estimated by a conventional observer, using pressure measurements. One possible solution to this estimation problem was presented in \protect\cite{dallabonaModelfreeSlidingMode2024}.

\subsection{Standard control input derivation}

Directly isolating $u$ in \eqref{eq:vd} is not possible: while it may be possible to invert the $K(u)$ function in \eqref{eq:flows}, the control input also appears inside the Heaviside functions, generating an algebraic loop where the derived input depends on the input itself. The standard way to solve this problem was  described by \cite{ayalewCascadeTuningNonlinear2006},:
\begin{equation}
\begin{aligned} \label{Q_p}
    Q_p(t,u) = & K(u)\left[\sqrt{(p_s-p_p)\,H(u)}+ \sqrt{(p_p-p_t)\,H(-u)}\right]\\
    \approx & Ku \left[\sqrt{(p_s-p_p)\,H(u_{-1})}+ \sqrt{(p_p-p_t)\,H(-u_{-1})}\right]\\
    = &K_{Qp}\,u,
\end{aligned}
\end{equation}

\begin{equation}
\begin{aligned} \label{Q_r}
    Q_r(t,u) = & K(u)\left[\sqrt{(p_r-p_t)\,H(u)}+ \sqrt{(p_s-p_r)\,H(-u)}\right]\\
    \approx & Ku \left[\sqrt{(p_r-p_t)\,H(u_{-1})}+ \sqrt{(p_s-p_r)\,H(-u_{-1})}\right]\\
    = &K_{Qr}\,u,
\end{aligned}
\end{equation}
where the term $u_{-1}$ represents the value of the input at the previous time step. In this way, the algebraic loop is eliminated and $u$ enters linearly in both equations. By defining
\begin{equation}
g=\frac{A_p B}{M_{} V_p} K_{Qp}+\frac{A_rB}{M_{}V_r} K_{Qr},
\end{equation}
a standard feedback linearizing control is obtained by,
\begin{equation}
u=\frac{1}{g} \left[ \frac{\dot{\hat{F}}_{d}}{M} +\frac{A_p^2 B}{MV_p}v_c + \frac{A_r^2 B}{MV_r}v_c + r^{(3)} - K_{FL}\underline{\mathbf{e}}\right].
\end{equation}

The $K(u)$ linearization is then reflected into uncertainties in the input term $g$ and in the terms $K_{Qp}$ and $K_{Qr}$ from  \eqref{Q_p} and \eqref{Q_r}.

\section{Cascaded control design}\label{sec:virt_inp}

\begin{figure}[h!]
    \centering
    \includegraphics[width=0.98\columnwidth]{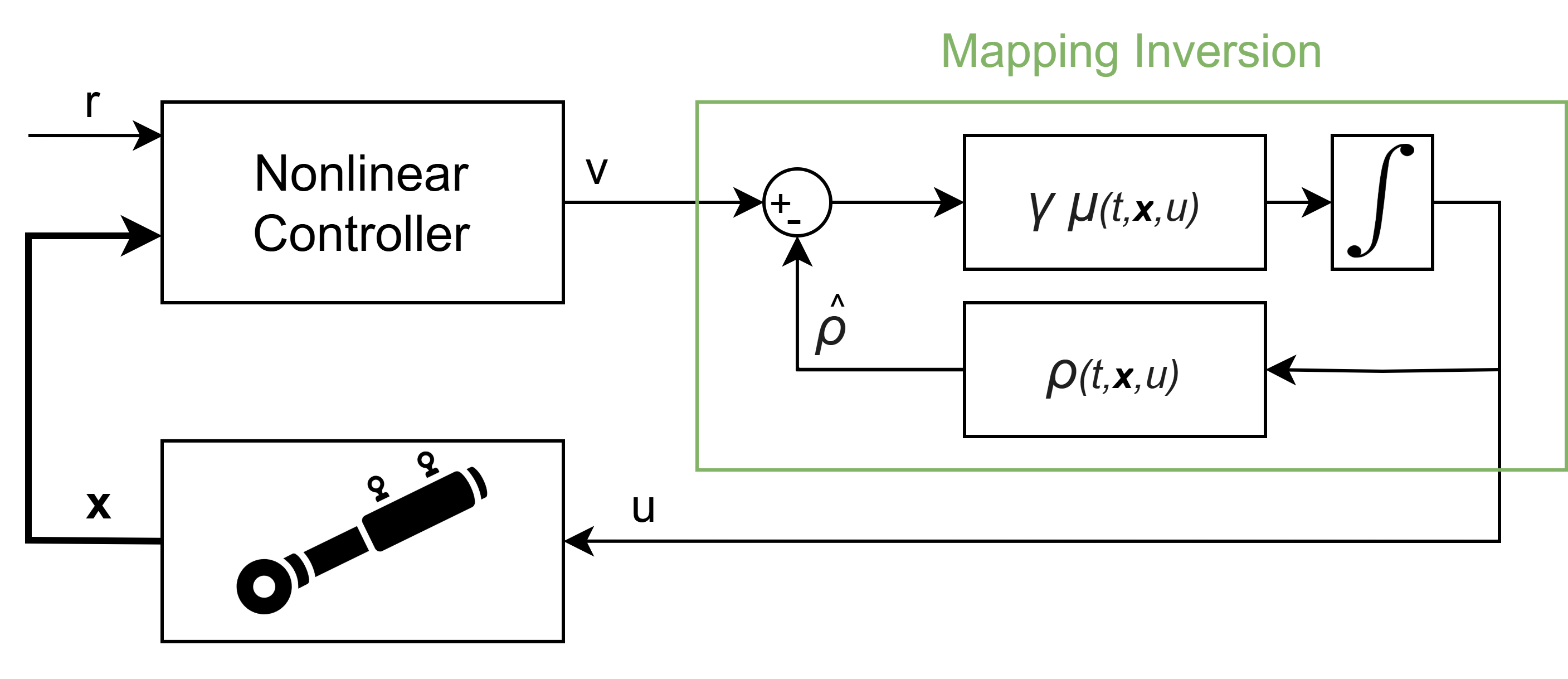}
    \caption{Block diagram of the proposed algorithm: Cascaded control with dynamic input mapping inversion. The input inversion is explained below.}
    \label{fig:block_diag}
\end{figure}

The delay operator introduced to avoid the formation of algebraic loops simplified the controller implementation but have the drawback of increasing unwanted chatter on the control signal \cite{ayalewCascadeTuningNonlinear2006},with associated consequences on oil quality degradation that were mentioned above. The solution proposed in this article, to avoid this, is to leverage a virtual input for the controller design and dynamically deriving a real input that produces the virtual one required by the controller. The block diagram for the proposed algorithm is shown in Fig.~\ref{fig:block_diag}.

Defining a virtual input
\begin{equation}\label{eq:virt_inp}
    v \triangleq \frac{A_pB}{MV_p}Q_p(u) + \frac{A_rB}{MV_r}Q_r(u),
\end{equation}
\eqref{eq:ddde} can be rewritten as
\begin{equation}\label{eq:ddde_v}
e^{(3)} =  -\frac{\dot{F}_{fric} + \dot{F}_{ext}}{M} - \frac{A_p^2B}{MV_p}v_c - \frac{A_r^2B}{MV_r}v_c - r^{(3)} + v,
\end{equation}
where $v$ is a function of time and the control input $u$, and is now present as a linear term. A feedback linearizing controller is now  defined in terms of an associated virtual input as
\begin{equation}\label{eq:v_fl}
v=\frac{\dot{\hat{F}}_{d}}{M} + \frac{A_p^2 B}{MV_p}v_c + \frac{A_r^2 B}{MV_r}v_c + r^{(3)} - K_{FL}\underline{\mathbf{e}}.
\end{equation}

\subsection{Dynamic mapping inversion algorithm}

The last part of the proposed method constitutes the derivation of the control input by solving \eqref{eq:virt_inp} for $u$. A dynamic nonlinear mapping inversion scheme is adopted from \cite{gripParameterEstimationCompensation2010b} for this purpose, allowing for re-formulating of the input calculation into a parameter estimation problem in $u$ for a nonlinearly parametrized regressor.

Specifically, let $\rho:[0,\infty)\times \mathbb{R}^3\times [-1,1] \rightarrow \mathbb{R}$ with
\begin{equation}\label{eq:rho_definition}
    \rho(t,\boldsymbol{x},u) = \frac{A_pB}{MV_{p}}Q_p(t,\boldsymbol{x},u) +  \frac{A_rB}{MV_{r}}Q_{r}(t,\boldsymbol{x},u)
\end{equation}

be a mapping from any given actual control input $u$ to the virtual input that was defined in \eqref{eq:virt_inp}. Assume that for every value $v$ at each time instant there exists an appropriate value of the control input $u^{*}$ such that $v = \rho(u^{*})$. It has been shown \cite{gripParameterEstimationCompensation2010b,papageorgiouRobustBacklashEstimation2019} that the estimation law
\begin{equation}\label{eq:estimation_law}
    \dot{u} = \gamma\,\mu(t,\boldsymbol{x},u) \, \Big( v - \rho(t,\boldsymbol{x},u) \Big),
\end{equation}
where $\gamma > 0$ is a tuning gain and $\mu(t,\boldsymbol{x},u)$ a function to be designed, ensures exponential convergence of the estimation error $\tilde{u} \triangleq u^{*} - u$ to the origin if the following conditions hold for all pairs $(u^{*},u)$:
\begin{condition}\label{con1} \normalfont
    There exists a positive definite real function $\sigma(t,\boldsymbol{x})$ such that
    \begin{equation}
    	\mu(t,\boldsymbol{x},u^{*})\left.\frac{\partial \rho}{\partial u}\right|_u + \left. \frac{\partial \rho}{\partial u}\right|_{u^{*}} \mu(t,\boldsymbol{x},u) \geq 2\sigma(t,\boldsymbol{x}) ~.
    \end{equation}
\end{condition}
\begin{condition}\label{con2} \normalfont
    There exists a positive real number $L_{\mu} > 0$ such that
    \begin{equation}
    	\left \vert \rho(t,\boldsymbol{x},u^{*}) - \rho(t,\boldsymbol{x},u) \right \vert \leq L_{\mu} \vert \tilde{u} \vert \sqrt{\sigma(t,\boldsymbol{x})} \; .
    \end{equation}
\end{condition}
\begin{condition}[Persistency of excitation]\label{con3} \normalfont
    There exist positive constants $T,\alpha_0 > 0$ such that
    \begin{equation}
    	\int_t^{t + T} \sigma(\tau,\boldsymbol{x}(\tau)) d\tau \geq \alpha_0 \boldsymbol{I} \; ,
     \end{equation}
     where $\boldsymbol{I}$ is the identity matrix.
\end{condition}

\begin{figure}[t]
    \centering
    \includegraphics[width=0.88\columnwidth]{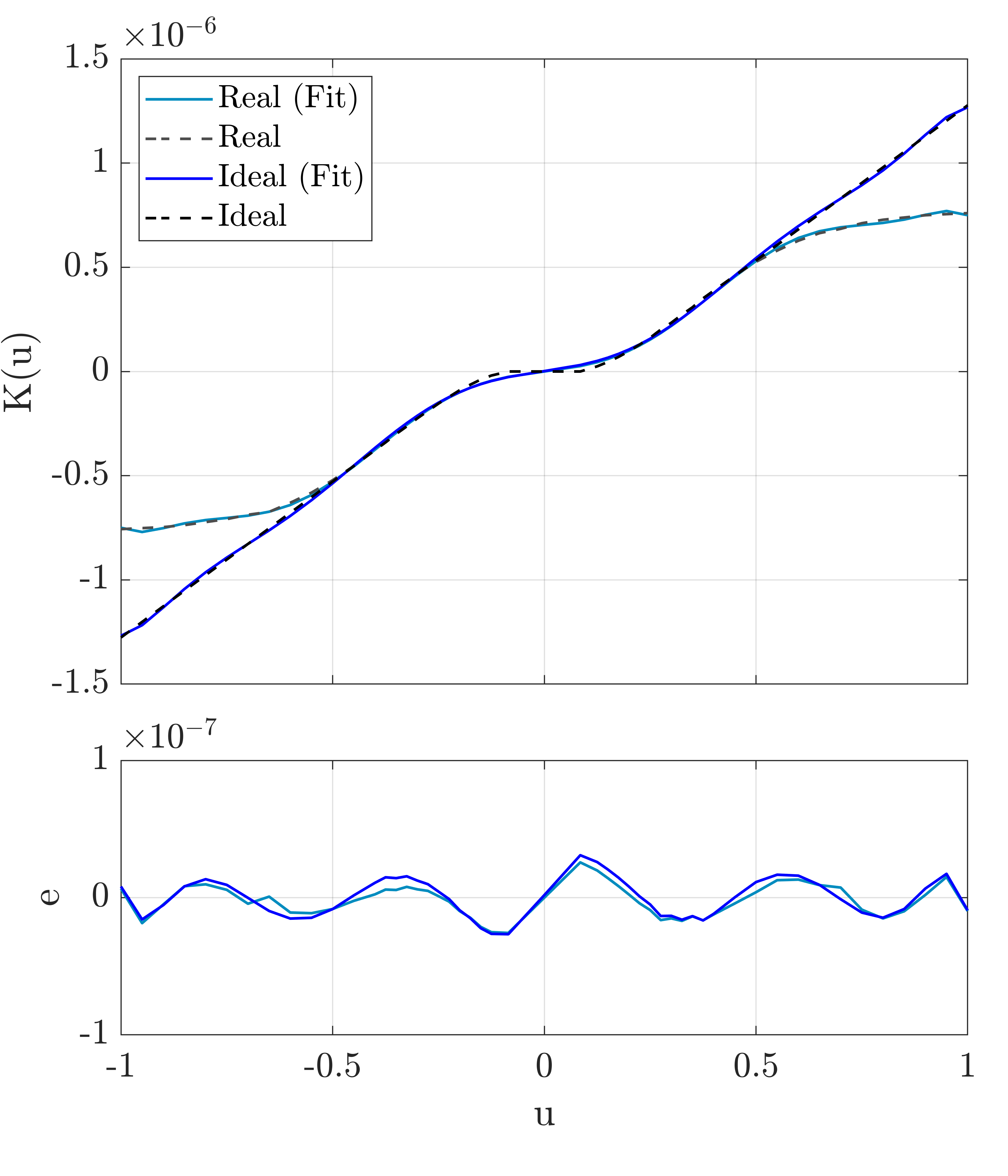}
    \caption{Manufacturer's lookup table of valve characteristic and approximation with a polynomial function.}
    \label{fig:mu_design}
\end{figure}

The final control action is then given by
\begin{equation}\label{eq:final_u}
    u(t) = \gamma \int_{0}^{t} \mu\big(\boldsymbol{x}(\tau),u(\tau)\big) \, \Big( v(\tau) - \rho\big(\boldsymbol{x}(\tau),u(\tau)\big) \Big) d\tau,
\end{equation}
\hl{where $v$ is the one defined in} \eqref{eq:v_fl}.

\begin{remark} \normalfont
    If the mapping $\rho$ is not injective, then under Conditions \ref{con1}-\ref{con3}, the estimation law in \eqref{eq:estimation_law} will ensure that $u$ converges to an appropriate value such that $\rho(t,u)$ converges to $v$ exponentially fast.
\end{remark}

\begin{figure}[t]
\centering\includegraphics[width=0.93\columnwidth]{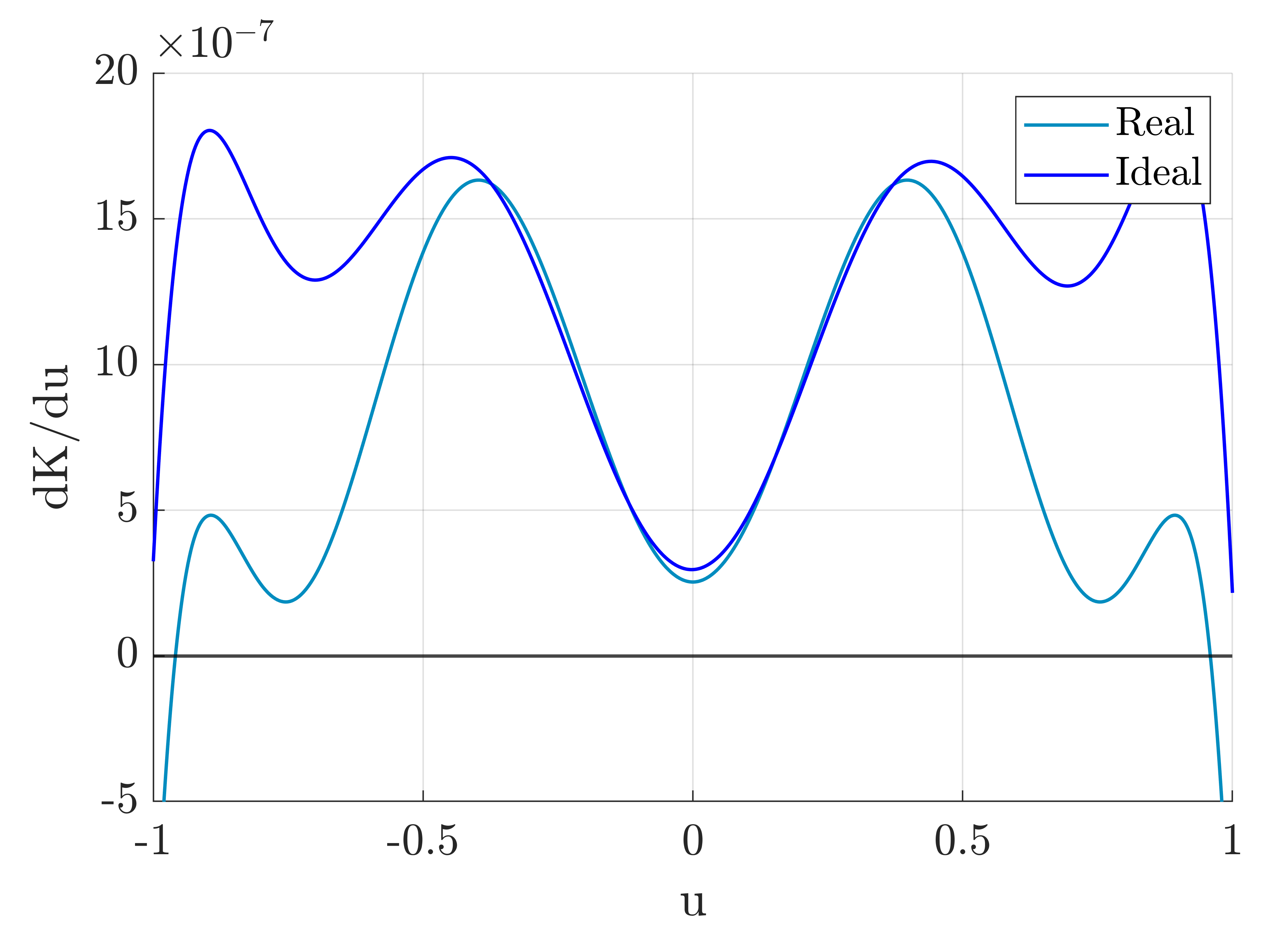}
    \caption{Flow coefficient derivative for algorithm convergence condition and valve degradation effect.}
    \label{fig:mu_check}
\end{figure}

A typical choice for $\mu$ is the gradient of $\rho$ with respect to $u$. However, computation of $\mu$ requires differentiation of the nonlinear function $K(u)$ present in \eqref{eq:flows} and, since the latter is usually available as a lookup table, it is approximated by an appropriate polynomial function of $u$.
Fig.~\ref{fig:mu_check} illustrates a ninth-order polynomial approximation for $K(u)$, i.e.
\begin{equation}\label{eq:poly_fit}
    K(u) = \sum_{i=0}^{9} \alpha_i \,u^i,
\end{equation}
for both an ideal actuator and the deteriorated (denoted \emph{real} in Fig 5) actuator used later in the  experimental validation.

Such approximation also ensures that for the case of the ideal valve $\displaystyle \frac{\partial K}{\partial u} >~0, \; \forall u\in[-1,1]$, as shown in Fig.~\ref{fig:mu_check}. This condition will be useful later in proving the validity of the convergence conditions. The function of $\mu$ can now be selected as
\begin{equation}\label{eq:d_mu}
    \mu(t,\boldsymbol{x},u) \triangleq \frac{\partial K}{\partial u}\left(\frac{A_pB}{MV_{p}}\sqrt{\Delta_{pp}(u)} + \frac{A_rB}{MV_{r}}\sqrt{\Delta_{pr}(u)}\right) > 0,
\end{equation}
since the quantity in the parenthesis is always positive. Evaluation of Condition \ref{con1} leads to
\begin{equation}
\begin{aligned}
\mu (t,\boldsymbol{x},u^{*}) \left. \frac{\partial \rho}{\partial u} \right|_{u} &+ \mu (t,\boldsymbol{x},u) \left. \frac{\partial \rho}{\partial u} \right|_{u^{*}} = 2 \mu (t,\boldsymbol{x},u^{*}) \mu (t,\boldsymbol{x},u)\\
&\geq 2 \left( \inf\limits_u \mu(t,\boldsymbol{x},u) \right)^2 \triangleq \sigma(t,\boldsymbol{x}) > 0 \; .
\end{aligned}
\end{equation}
Moreover, since $\left \vert \displaystyle \frac{\partial \rho}{\partial u} \right \vert$ is bounded, there exists a Lipschitz constant $L_u > 0$ such that
\begin{align}
    \left \vert \rho(t,\boldsymbol{x},u^{*}) - \rho(t,\boldsymbol{x},u) \right \vert < L_u \vert \tilde{u} \vert \leq \frac{L_u \vert \tilde{u} \vert \sqrt{\sigma(t,\boldsymbol{x})}}{\sqrt{\inf\limits_{(t,\boldsymbol{x})}\sigma(t,\boldsymbol{x})}}
\end{align}
which fulfills Conditions \ref{con1},\ref{con2} with
$$
    L_{\mu} = \displaystyle \frac{L_u}{\sqrt{\inf\limits_{(t,\boldsymbol{x})}\sigma(t,\boldsymbol{x})}} \; .
$$
Finally, Condition \ref{con3} is satisfied by the definition of $\sigma$ since $\sigma(t,\boldsymbol{x}) > 0$, $\forall (t,\boldsymbol{x})\in\mathbb{R}_{\geq 0}\times \mathbb{R}^3$.

\begin{remark}
The $\mu$ function that was chosen for the purpose is an approximation of the gradient, where the dependency on u in the terms $\Delta_{pp}$ and $\Delta_{pr}$ is neglected during differentiation, since it only appears inside a Heaviside function. Hence, it would only be reflected in additional spikes in case a discontinuity is present during switching.
\end{remark}

\begin{remark}
    In the case of a non-ideal valve, e.g. due to wear, the term $\displaystyle\frac{\partial K}{\partial u}$ may take non-positive values as illustrated in Fig.~\ref{fig:mu_check}, which may lead to violation of Condition~\ref{con1}. This will, in turn, result to a bounded input estimation error. An alternative yet more conservative choice is to select $\mu$ to be a positive constant.
\end{remark}

\section{Stability Analysis}\label{sec:stab}

\begin{figure}[t]    \centering\includegraphics[width=0.95\columnwidth]{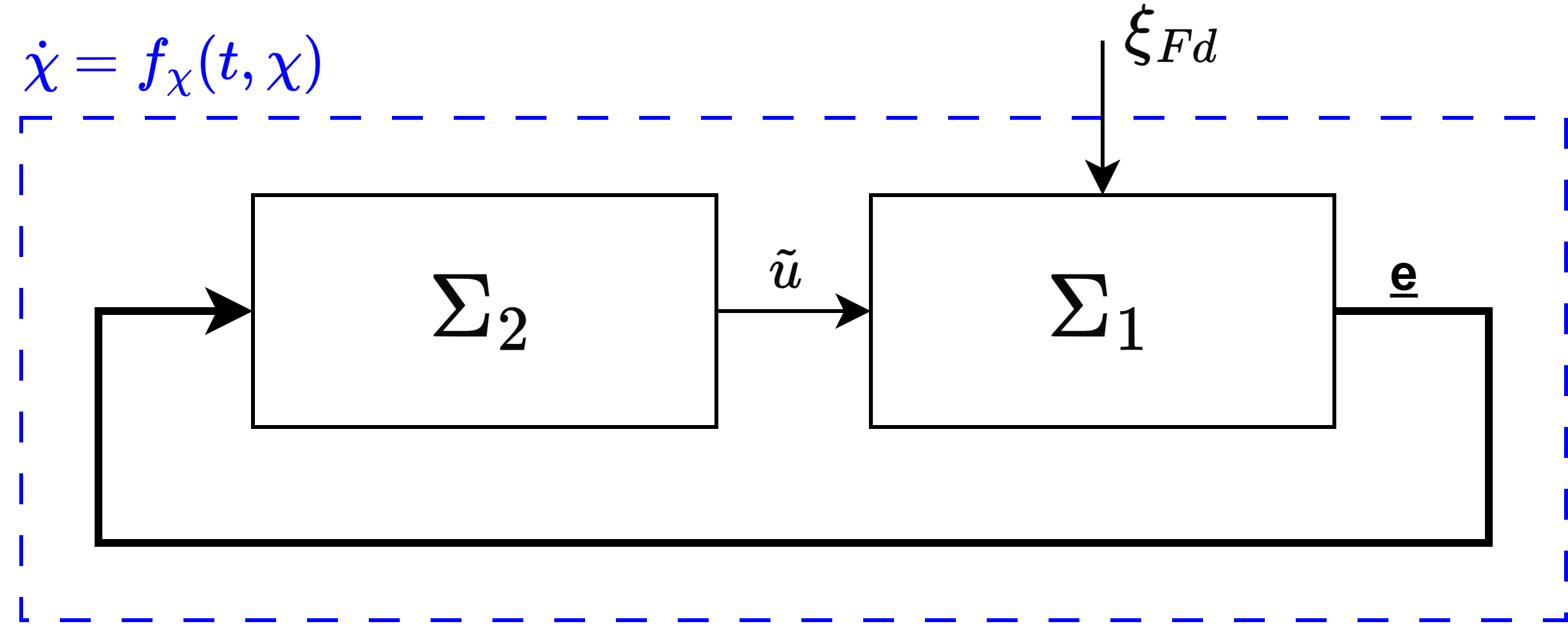}
        \caption{\hl{Closed loop system $\boldsymbol{\dot{\chi}} = \boldsymbol{f_{\chi}}(t,\boldsymbol{\chi}) + \boldsymbol{B_{\chi}}\boldsymbol{\xi_{F_d}}$.}}
        \label{fig:ClosedLoopSysteM}
    \end{figure}

The stability of the closed-loop system is analyzed by including the nonlinear controller, the dynamical mapping inversion, and the disturbance estimator.
Let $\delta_{\underline{\mathbf{e}}}$ be the error associated with the calculation of $\ddot{e}$. The dynamics of the closed-loop tracking error $\underline{\mathbf{e}}$ is written as:

    \begin{align}
        \dot{\underline{\mathbf{e}}} =& \boldsymbol{A_{FL}}\underline{\mathbf{e}} + \boldsymbol{B_{FL}}v = \boldsymbol{A_{FL}}\underline{\mathbf{e}} + \boldsymbol{B_{FL}}(v^*-\tilde{v}) \nonumber\\
        =& \boldsymbol{A_{FL}}\underline{\mathbf{e}} + \boldsymbol{B_{FL}}\left(-\boldsymbol{K_{FL}}(\underline{\mathbf{e}}-\delta_{\underline{\mathbf{e}}})-\frac{\dot{\tilde{F}}_d}{M} -\tilde{v}\right) \nonumber\\
        =& (\boldsymbol{A_{FL}}-\boldsymbol{B_{FL}K_{fl}})\,\underline{\mathbf{e}} + \boldsymbol{B_{FL}}\left(\boldsymbol{K_{FL}}\begin{bmatrix}
        0\\ -\frac{\tilde{F}_{d}}{M}
    \end{bmatrix}-\frac{\dot{\tilde{F}}_d}{M}\right) \nonumber\\
    &- \boldsymbol{B_{FL}}\,\left[\rho(u^*)-\rho(u^*-\tilde{u})\right] \nonumber\\
    =& (\boldsymbol{A_{FL}}-\boldsymbol{B_{FL}K_{fl}})\,\underline{\mathbf{e}} + \boldsymbol{\xi_{F_d}} + \boldsymbol{h_{FL}}(t,\tilde{u}).
    \end{align}
with
\begin{align}
    &\boldsymbol{\xi}_{F_d} = \boldsymbol{B_{FL}} \left( -\frac{\dot{\tilde{F}}_{d}}{M} + \boldsymbol{K_{FL}}\begin{bmatrix}
        0\\ -\displaystyle \frac{\tilde{F}_{d}}{M}
    \end{bmatrix}\right)\\
    &\boldsymbol{h_{FL}}(t,\tilde{u}) =  -\boldsymbol{B_{FL}}\,\left[\rho(u^*)-\rho(u^*-\tilde{u})\right],
\end{align}
Moreover, let $v^* = \rho(u^*)$ be the required virtual controller and $u^*$ the related control input. Then, one can define
\begin{equation}
        \tilde{v} \triangleq v^*-v = \rho(u^*) - \rho(u) = \rho(u^*) - \rho (u^* - \tilde{u}),
\end{equation}
 for some arbitrary $u$ and $v$. From \eqref{eq:estimation_law} the dynamics of the input estimation error $\tilde{u}$ reads:
\begin{equation} \label{eq:u_tilde_dynamics}
    \dot{\tilde{u}} = \dot{u}^* - \gamma \, \mu(x, u^*, \tilde{u})\, \Big( -\boldsymbol{K_{FL}}\underline{\mathbf{e}} - \rho (u^* - \tilde{u}) \Big) \triangleq f_2(t,\underline{\mathbf{e}},\tilde{u})
\end{equation}
The entire closed-loop system can therefore be \hl{written as}
\begin{equation} \label{eq:entire_closed_loop_system}
    \underbrace{\begin{bmatrix}
        \dot{\underline{\mathbf{e}}}\\
        \dot{\tilde{u}}
    \end{bmatrix}}_{\boldsymbol{\dot{\chi}}} = \underbrace{\begin{bmatrix}
        (\boldsymbol{A_{FL}}-\boldsymbol{B_{FL}K_{fl}})\,\underline{\mathbf{e}} + \boldsymbol{h_{FL}}(t,\tilde{u})\\
        f_2(t,\underline{\mathbf{e}},\tilde{u})
    \end{bmatrix}}_{\boldsymbol{f_{\chi}}(t,\boldsymbol{\chi})} + \underbrace{\begin{bmatrix}
        \boldsymbol{I_3}\\
        \boldsymbol{0_{1\times 3}}
    \end{bmatrix}}_{\boldsymbol{B_{\chi}}}\boldsymbol{\xi_{F_d}}.
\end{equation}
\begin{lemma} \label{lem:unperturbed_closed_loop}
    \hl{The unperturbed closed-loop system $\boldsymbol{\dot{\chi}} = \boldsymbol{f_{\chi}}(t,\boldsymbol{\chi})$ ($\boldsymbol{\xi_{F_d}} = \boldsymbol{0}$) is Uniformly Globally Asymptotically Stable (UGAS).}
\end{lemma}
\begin{proof}
    The system can be seen as the feedback interconnection of the following subsystems:
    \begin{equation}\label{eq:inter_sys}
        \begin{aligned}
            &\Sigma_1: \dot{\underline{\mathbf{e}}} = (\boldsymbol{A_{FL}}-\boldsymbol{B_{FL}K_{fl}})\,\underline{\mathbf{e}} + \boldsymbol{h_{FL}}(t,\tilde{u})\\
            &\Sigma_2: \dot{\tilde{u}} = f_2(t,\underline{\mathbf{e}},\tilde{u})
        \end{aligned}
    \end{equation}
as illustrated in Fig. \ref{fig:ClosedLoopSysteM}.
    
    The following statements are true:
    \begin{enumerate}[label=\textbf{S\arabic*}:]
        \item The unperturbed subsystem $\Sigma_1$ $(\tilde{u}=0)$ is Uniformly GES (UGES).
        \item The origin of the subsystem $\Sigma_2$ is UGES.
        \item There exist two scalar, positive-definite functions $V_{FL},W$ such that
        \begin{equation}
            \frac{\partial V_{FL}(t, \underline{\mathbf{e}})}{\partial \underline{\mathbf{e}}} \boldsymbol{h_{FL}}(t, \tilde{u}) = o(W(\underline{\mathbf{e}})), \; \forall (t,\underline{\mathbf{e}}) \in [0,\infty) \times \mathbf{R}^3 .
        \end{equation}
    \end{enumerate}
    The validity of \textbf{S1} directly follows from the fact that $\boldsymbol{A_{FL}}-\boldsymbol{B_{FL}K_{fl}}$ is Hurwitz by design and hence, with $\tilde{u} = 0$, the system $\dot{\underline{\mathbf{e}}} = (\boldsymbol{A_{FL}}-\boldsymbol{B_{FL}K_{fl}})\,\underline{\mathbf{e}}$ has a UGES equilibrium at origin. The statement \textbf{S2} is verified by the design of the estimator in \eqref{eq:estimation_law}, which under Conditions 1-3, guarantees exponential stability of the origin of the estimation error $\tilde{u}(t)$ at all times. Finally, consider the scalar, positive-definite Lyapunov function $V_{FL} = \underline{\mathbf{e}}^TP_{FL}\,\underline{\mathbf{e}}$, where $P_{FL}$ is the symmetric, positive-definite solution to the Lyapunov matrix equation
$$
    (\boldsymbol{P_{FL}A_{FL}}-\boldsymbol{B_{FL}K_{fl}}) + (\boldsymbol{A_{FL}}-\boldsymbol{B_{FL}K_{fl}})^T\boldsymbol{P_{FL}} = -\boldsymbol{I}
$$
associated to the UGES origin of the unperturbed $\Sigma_1$. Then 
    \begin{align}
        &\left\| \frac{\partial V_{FL}}{\partial \underline{\mathbf{e}}} \boldsymbol{h_{FL}}(t, \tilde{u}) \right\| \leq \left\| 2\boldsymbol{P_{FL}}\underline{\mathbf{e}} \right\| \,\left\| \boldsymbol{B_{FL}} \right\|\, \left\vert \rho(u^*) - \rho(u^* - \tilde{u}) \right\vert \nonumber\\
        &\leq \left\| 2\boldsymbol{P_{FL}}\underline{\mathbf{e}} \right\| \, \left\| \boldsymbol{B_{FL}} \right\| \, L_{u} \, \vert \tilde{u} \vert \leq \underbrace{\left\| 2\boldsymbol{P_{FL}} \right\| \,\left\| \boldsymbol{B_{FL}} \right\| \, L_{u} \, \vert \tilde{u} \vert_{\infty}}_{\eta > 0}\, \left\| \underline{\mathbf{e}} \right\| \nonumber\\
        &= \eta \Vert \underline{\mathbf{e}} \Vert^2 \triangleq W(\Vert \underline{\mathbf{e}} \Vert). \label{ineq:gradient}
    \end{align}
    The last inequality implies $\displaystyle \frac{\partial V_{FL}(t, \underline{\mathbf{e}})}{\partial \underline{\mathbf{e}}} \boldsymbol{h_{FL}}(t, \tilde{u}) = o(W(\underline{\mathbf{e}}))$, $\forall (t,\underline{\mathbf{e}}) \in [0,\infty) \times \mathbf{R}^3$.

Following the approach introduced in \cite{loriaFeedbackCascadeinterconnectedSystems2008}, \hl{the unperturbed closed-loop system }\eqref{eq:inter_sys} can be seen as a cascaded interconnection of $\Sigma_1$ and $\Sigma_2$, where the solutions $\underline{\mathbf{e}}(t;\underline{\mathbf{e}}_{\boldsymbol{0}})$ of $\Sigma_1$ for all initial conditions $\underline{\mathbf{e}}_{\boldsymbol{0}}$ can be thought of as \emph{time-varying parameters} of $\Sigma_2$. Then since statements \textbf{S1}, \textbf{S2}, \textbf{S3} are true, by \hl{Proposition} 1 in  \cite{loriaFeedbackCascadeinterconnectedSystems2008} the solutions of the \hl{unperturbed} closed-loop system are globally \hl{UGAS}.
\end{proof}

\begin{prop}
    \hl{The solutions of the entire closed-loop system in Eq. }\eqref{eq:entire_closed_loop_system}\hl{ are Uniformly Ultimately Bounded (UUB).}
\end{prop}
\begin{proof}
    \hl{This follows directly from Lemma }\ref{lem:unperturbed_closed_loop}\hl{ and Lemma 9.3 in }\cite{khalilNonlinearSystems2002}\hl{ since the bounded perturbation $\boldsymbol{\xi_{F_d}}$ is non-vanishing at the origin.}
\end{proof}

\section{Case Study}\label{sec:case_study}

\begin{figure}[tp]
    \centering
    \includegraphics[width=0.92\columnwidth]{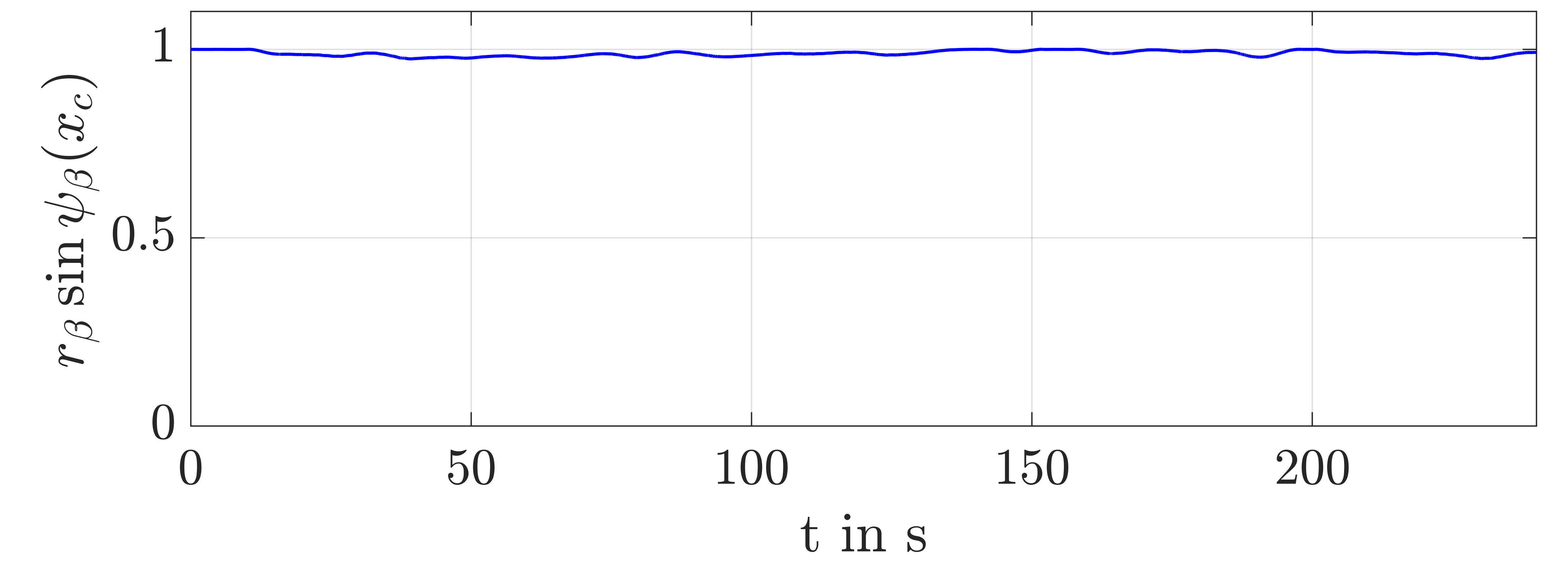}
    \caption{Momentum arm angle for verifying that constant assumption holds.}
    \label{fig:psi_const}
\end{figure}

The case study on which the algorithm is presented is a wind turbine pitch system with the hydraulic actuator used to provide a time-varying blade pitch, mainly according to prevailing wind and power demand. The mechanical link between cylinder and turbine blade is stiff, such that controlling the blade angle is equivalent to controlling cylinder position. The angular dynamics of the blade is described by:
\begin{equation}\label{eq:pitch_dyn}
\begin{aligned}
& J_{tot} \dot{v_\beta}=\overbrace{\left(A_p p_p-A_r p_r\right) r_{\beta} \sin \left(\psi\left(x_c\right)\right)}^{M_h}-M_{f r i c}-M_{e x t} \\
& \dot{\beta}=v_\beta,
\end{aligned}
\end{equation}
where the equivalent mass $M$ now also accounts for blade inertia. \eqref{eq:pitch_dyn} matches the dynamics of the cylinder dynamics presented in \eqref{eq:final_model}. The presence of a cyclic component in the variable distance that relates mechanical moment to force is minor, as shown in Fig.~\ref{fig:psi_const}. Thus, the control design aims at tracking the angle while the cylinder dynamics represents a stable internal dynamics, under the assumption of having a stiff mechanical connection between cylinder and blade. Hence, there exists a static transformation relating $x_c$ to $\beta$ and with the pitch tracking error $e_{\beta}$ being the difference between pitch angle and a reference, 
\begin{equation}
e_{\beta}=\beta - r,
\end{equation}
the relative degree of the closed loop pitch control will be the same as for the closed loop cylinder piston position control loop analyzed above. With relative degree three of both, the third derivative of the new tracking error is needed,
\begin{align}
e_{\beta}^{(3)} = & -\frac{\dot{M}_{fric}+ \dot{M}_{ext}}{J_{tot}}\\
    &- \frac{r_{\beta} A_p^2B}{J_{tot}V_p}v_c\sin{\psi_{\beta}(x_c)} - \frac{r_{\beta} A_r^2B}{J_{tot}V_r}v_c\sin{\psi_{\beta}(x_c)} \notag \\
    & + \frac{r_{\beta} A_pB}{J_{tot}V_p}Q_p(u)\sin{\psi_{\beta}(x_c)} + \frac{r_{\beta} A_rB}{J_{tot}V_r}Q_r(u)\sin{\psi_{\beta}(x_c)} \notag \\
    & + \underbrace{\frac{r_{\beta}}{J_{tot}} F_h \cos(\psi_{\beta}(x_c)) \frac{d\psi}{dx_c} v_c}_{\tau(x)}
     - r^{(3)} \, \triangleq  \, v.
\end{align}

With respect to the general case presented in \eqref{eq:ddde}, two variations are present:
\begin{itemize}
    \item $1/M$ is replaced by $r_{\beta}\sin{\psi_{\beta}(x_c)}/J_{tot}$.
    \item An additional term $\tau(x)$, is introduced.
\end{itemize}These changes arise due to the variable-length moment arm in the updated dynamics.

However, being $\psi_{\beta}(x_c)$ is approximately constant, the following assumptions are made:
\begin{itemize}
    \item $\psi_{\beta}(x_c) \approx \textit{const}$
    \item $\displaystyle \frac{d\psi_{\beta}}{dx_c} \approx 0 \rightarrow \tau(x) \approx 0$.
\end{itemize}

\subsection{Simulation Setup}

The designed control actions are \hl{tuned} on a high-fidelity simulator, which includes the general dynamics presented in \eqref{eq:gen_dyn} and \eqref{eq:pitch_dyn}, and also encompass the control valve dynamics. The measured signals are subjected to noise, simulated to match the levels in the test rig. The velocities ($v_c$ and $v_{\beta}$) are obtained via numerical differentiation since they are not usually available as sensor signals in a real plant. The reference for the pitch angle and the wind load $M_{ext}$ are obtained from a wind turbine high-fidelity simulator based on OpenFAST, widely used in wind turbine research and development \cite{Jonkman2013}.

\subsection{Experimental setup}

\begin{figure}[t]
\centering\includegraphics[width=0.8\columnwidth]{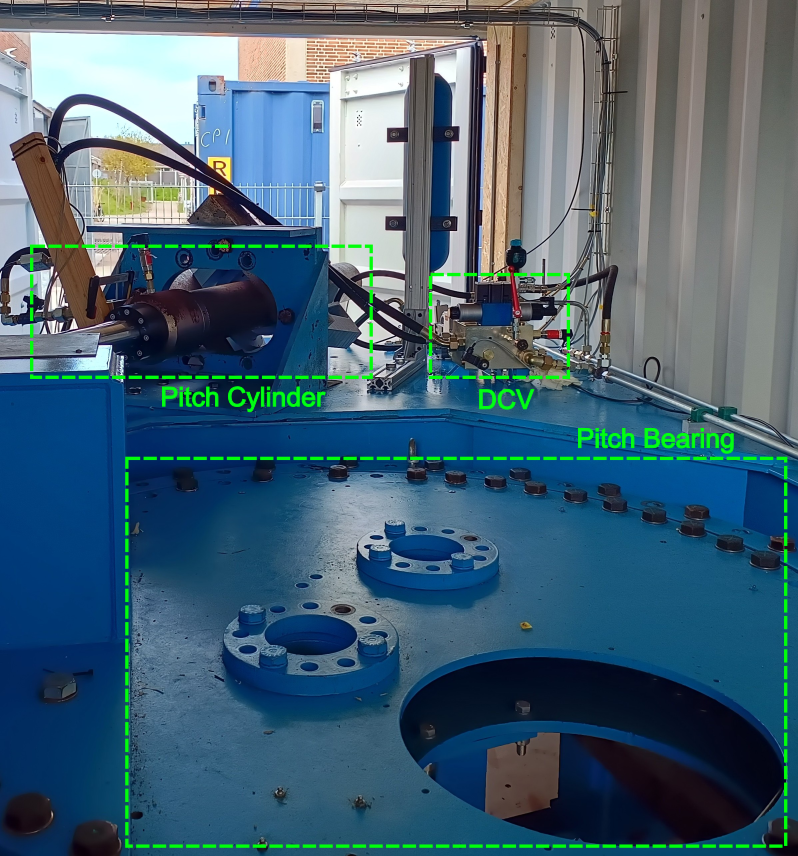}
    \caption{Full-scale experimental setup from a wind turbine: $104\,\mathrm{kg}$ hydraulic cylinder with a rated force of $150\,\mathrm{kN}$, blade mount and pitch bearing with an inertia of $280\,\mathrm{kg\,m}^2$. The 80 m blade is not attached, but its inertia is emulated through the action of a load cylinder, which excites the setup in real time, and also includes the OpenFAST calculated blade moment.}
    \label{fig:exp_pic}
\end{figure}

The proposed algorithm is further tested on a full scale pitch system from a real wind turbine with a rated power of 3 MW. A picture of the setup is presented in Fig. \ref{fig:exp_pic}, and further details can be found in \cite{pedersenOnlineLoadLeakage2025}.

\begin{figure}[t]
\centering\includegraphics[width=0.8\columnwidth]{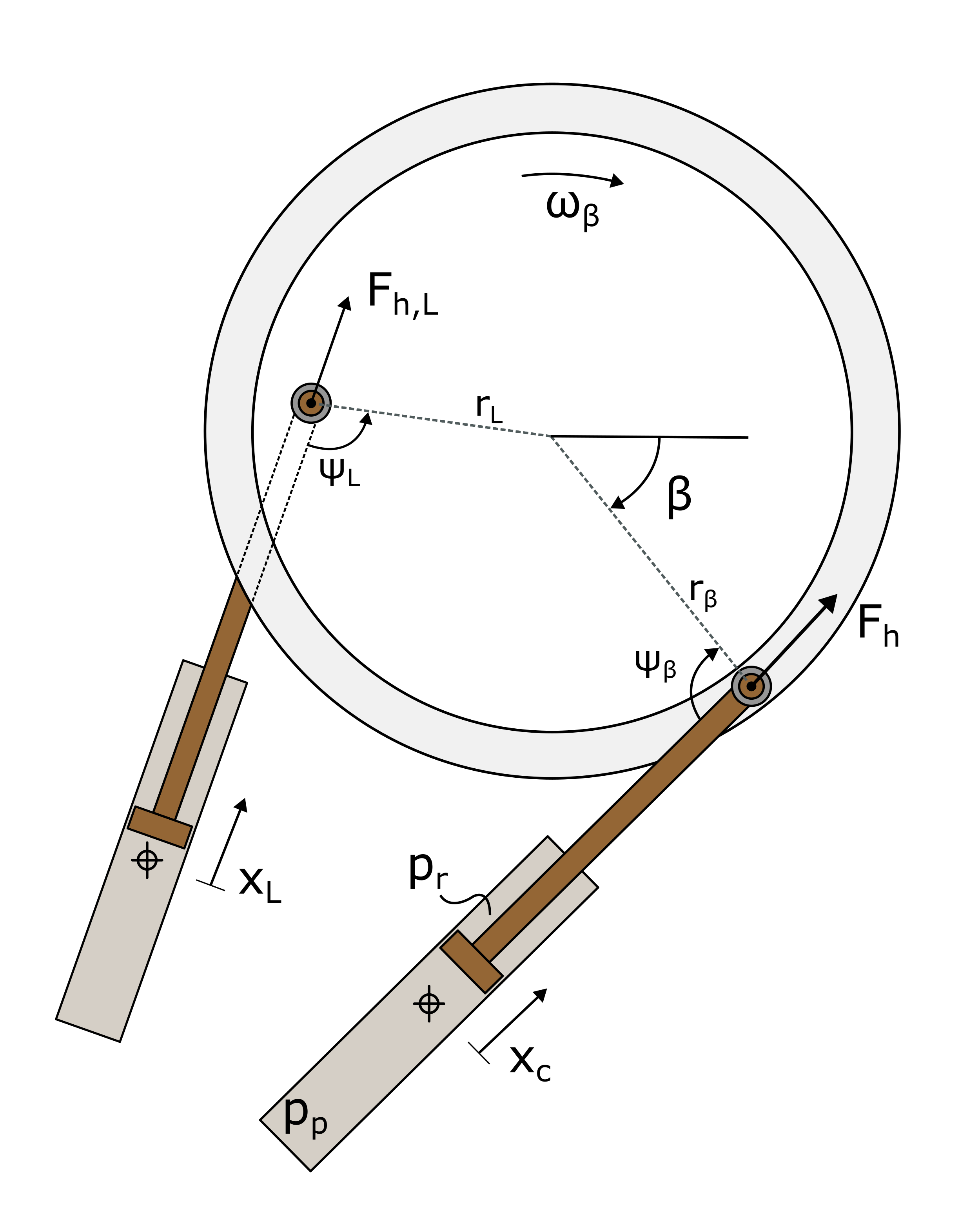}
    \caption{Schematics of the experimental setup with pitch control cylinder right, generating $F_h$, and load cylinder left, generating $F_{h,L}$ in real time.}
    \label{fig:exp_scheme}
\end{figure}

For practical reasons, the test rig does not include the wind turbine blade. However, an additional load cylinder is available to both replicate the wind load from OpenFAST and digitally reproduce the blade's inertia. Fig.~\ref{fig:exp_scheme} presents a detailed diagram of the test rig’s composition.

In the real turbine, the total inertia is given by the inertia of the pitch test rig ($J_{tr}$) plus the inertia of the blade ($J_{b}$). One can write:

\begin{equation}
    \begin{aligned}
        J_{tr}\dot{v_\beta} &= M_h-M_{f r i c}-M_{e x t} - J_{b}\dot{v_\beta}\\
        &\approx M_h-M_{f r i c}- \underbrace{\left(M_{e x t} + J_{b}\,\ddot{r} \right)}_{M_{h,l}},
    \end{aligned}
\end{equation}
by assuming $r \approx \beta $, and avoiding double differentiating the noisy position measurement. Thus, the inertia of the blade is reproduced by injecting an additional mechanical torque by means of the load cylinder.

The real-time implementation of the control action is performed using a Speedgoat real-time module for MATLAB/Simulink.

\subsection{Controllers Design}

The control action for the baseline feedback linearization controller and for the proposed algorithm are derived by adapting the methods described in Sec.~\ref{sec:standard_design} and Sec.~\ref{sec:virt_inp}, respectively.

\subsubsection{Standard controller}

The control action for the standard case is obtained by defining the input gain:
\begin{equation}
g=\frac{r_{\beta}A_p B}{J_{tot} V_p} K_{Qp}+\frac{r_{\beta}A_rB}{J_{tot}V_r} K_{Qr}.
\end{equation}
Finally, the feedback linearization controller is derived as,
\begin{equation}
\begin{aligned}
u&=\frac{1}{g} \left[ \frac{r_{\beta}A_p^2 B}{J_{tot}V_p}v_c + \frac{r_{\beta}A_r^2 B}{J_{tot}V_r}v_c + \frac{\dot{\hat{M}}_d/J_{tot} + r^{(3)}- K_{FL}\underline{\mathbf{e}}_{\beta}}{\sin({\psi_{\beta}(x_c)})} \right]\\
&\approx \frac{1}{g} \left[ \frac{r_{\beta}A_p^2 B}{J_{tot}V_p}v_c + \frac{r_{\beta}A_r^2 B}{J_{tot}V_r}v_c + \frac{\dot{\hat{M}}_d}{J_{tot}} + r^{(3)} - K_{FL}\underline{\mathbf{e}}_{\beta}\right]
\end{aligned},
\end{equation}
where $\hat{M_{d}}$ is available from the load cylinder, and the $\psi_{\beta}$ angle dependency is removed since it only affects the $r^{(3)}$ term and the term associated with the gain, which is a degree of freedom anyway. The vector $\underline{\mathbf{e}}_{\beta}$ contains the tracking error $e_{\beta}$ and its derivatives, similarly to \eqref{eq:fl_dynamics}.

\subsubsection{Mapping inversion based controller}

Following a similar approach, the virtual control input in the proposed solution is:
\begin{equation}
    v= \frac{\dot{\hat{M}}_d}{J_{tot}} +\frac{r_{\beta}A_p^2 B}{J_{tot}V_p}v_c + \frac{r_{\beta}A_r^2 B}{J_{tot}V_r}v_c +  r^{(3)} - K_{FL}\underline{\mathbf{e}}_{\beta},
\end{equation}
and the dynamical mapping inversion algorithm is 
\begin{equation}\label{eq:mi_cs}
\begin{aligned}
    &\dot{u} = \gamma \, \mu(t,\boldsymbol{x},u)  \big( v - \rho(t,\boldsymbol{x},u) \big)\\
    &\rho(t,\boldsymbol{x},u) = \frac{r_{\beta}A_pB}{J_{tot}V_{p}}Q_p(p_p,p_r,u) +  \frac{r_{\beta}A_rB}{J_{tot}V_{r}}Q_{rv}(p_r,u)\\
    &\mu(t,\boldsymbol{x},u) = \frac{\partial K}{\partial u}\left(\frac{A_pB}{MV_{p}}\sqrt{\Delta_{pp}(u)} + \frac{A_rB}{MV_{r}}\sqrt{\Delta_{pr}(u)}\right),
\end{aligned}
\end{equation}
with $K(u)$ defined according to \eqref{eq:poly_fit} for the case of the real deteriorated valve, and $u$ according to \eqref{eq:final_u}.

\section{Results}\label{sec:sim_res}

The controllers are tested with the same tuning both in simulation and in the experimental setup. Furthermore, the same gains are chosen for the baseline solution and the presented algorithm, with the latter having an additional tuning parameter $\gamma$ for regulating the adaptation rate in the dynamic inversion. Different solutions for different values for $\gamma$ are presented to analyze its impact.

\subsection{Tracking performance}

\begin{figure}[h!]
    \centering
    \includegraphics[width=1.02\columnwidth]{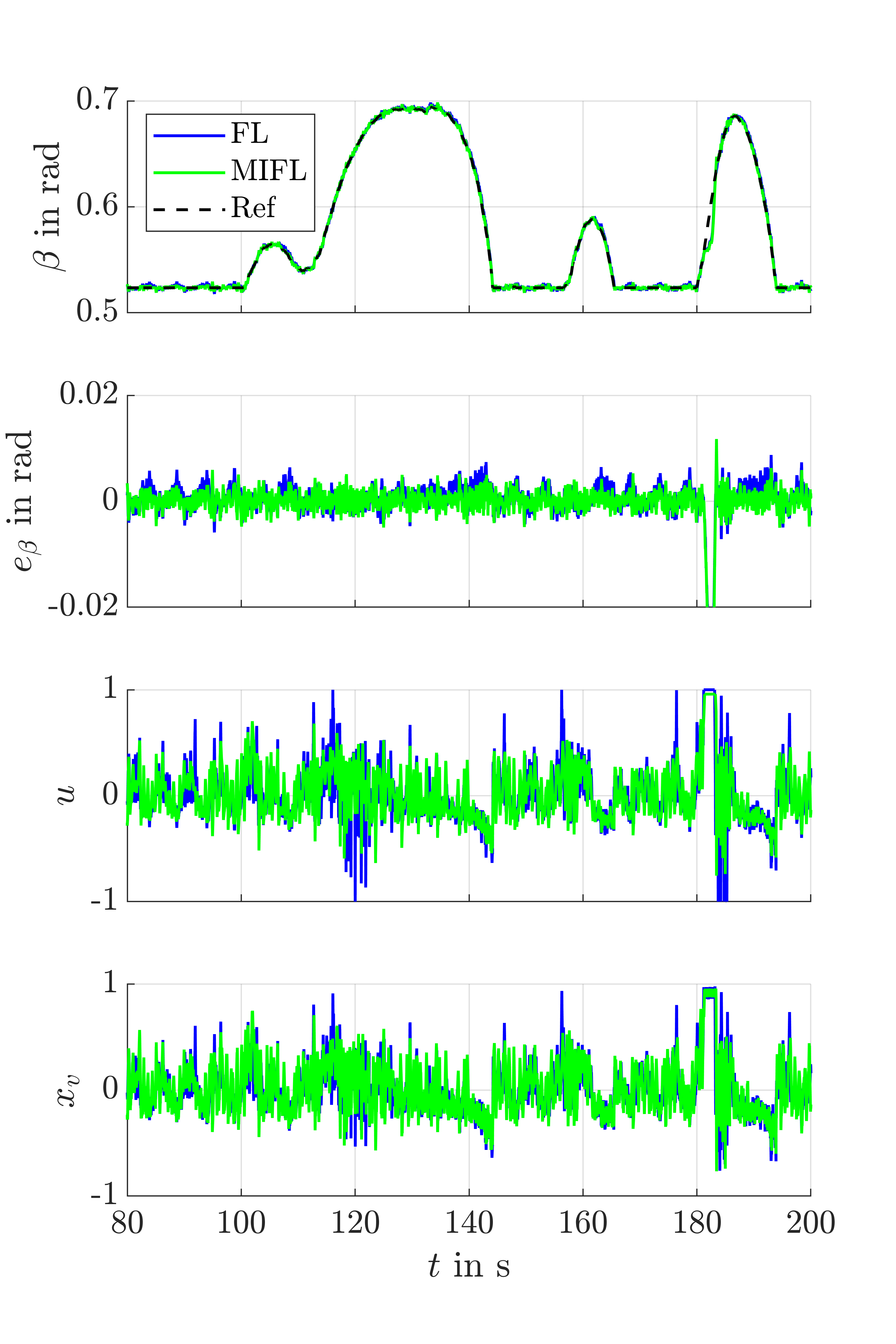}
    \caption{\hl{Experimental} tracking performance of the Feedback linearization based controller, with and without applying the dynamical mapping inversion.}
    \label{fig:fl_track}
\end{figure}

\begin{figure}[h!]
    \centering
\includegraphics[width=1.02\columnwidth]{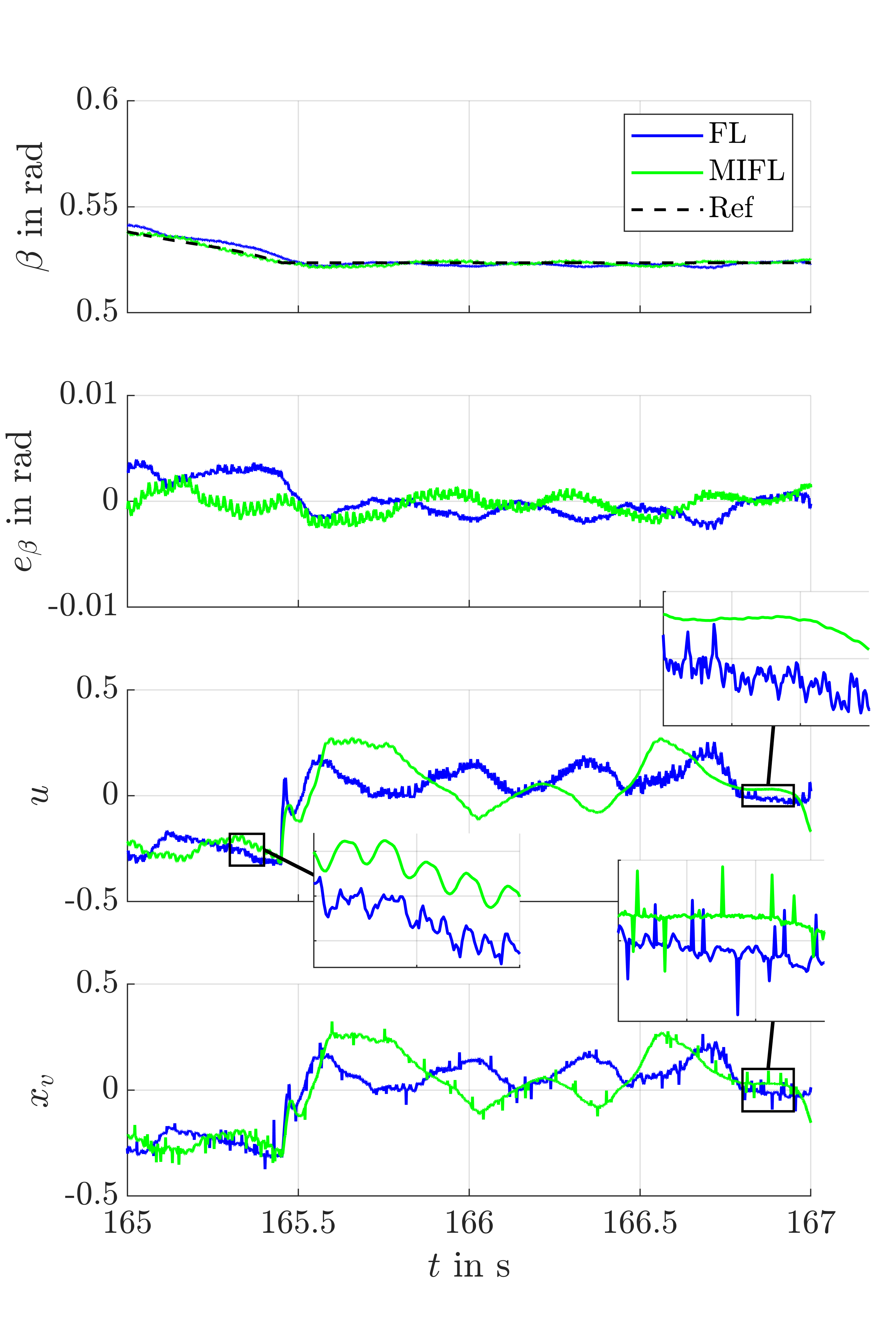}
    \caption{Zoomed on regions of interest for showing the tracking properties under different conditions.}
    \label{fig:fl_track_zoom}
\end{figure}

The experimental recordings for the two controllers are presented in Fig.~\protect\ref{fig:fl_track}, and in Fig.~\protect\ref{fig:fl_track_zoom} in a zoomed version.

Both the baseline FL and the proposed algorithm are able to track the reference with an error which is always smaller than 0.01 rad. Around 180 seconds tracking is temporary lost or both solutions, due to limitations in the common supply pressure shared by both hydraulic systems. Both controllers saturate and are then able to track the reference back as soon as the required force (hence flow) decreases for both. A quantitative comparison between the two solutions, discarding the interval of saturation, is presented later.

However, from the zoomed plot it is already possible to appreciate some differences between the two solutions. The control input $u$ displays reduced chattering, especially in the region where no pitching action is present. The control action appears smoother also in the region of active pitching, even if it also contains more similar oscillations with respect to the baseline solution. Such difference is smoothened when analyzing the actual spool's position $x_v$. In fact, the valve performs a natural filtering action according to its bandwidth. 

The signal quality in $u$ and $x_v$ are equivalent for the MI solution, while part of the noise is filtered out by the valve dynamics in the standard solution, since the controller doesn't have knowledge of the valve dynamics, nor of valve limitations.

\begin{remark}
    The spikes in the valve position can be originated by the internal closed-loop or by the internal position sensor.
\end{remark}

\subsection{Mapping inversion algorithm}\label{mi_track_res}

\begin{figure}[h!]
    \centering
\includegraphics[width=0.97\columnwidth]{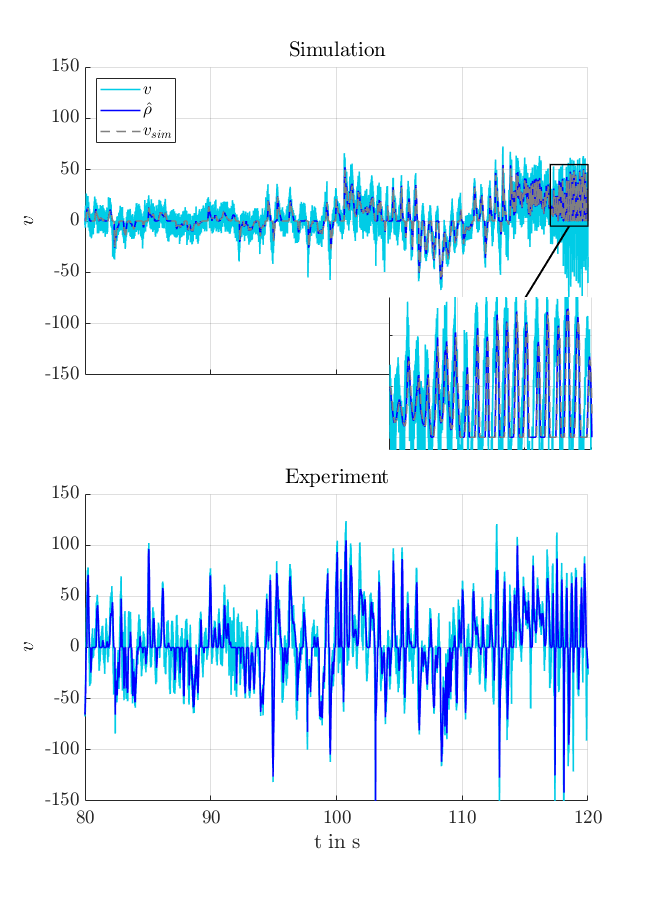}
    \caption{Dynamical mapping inversion algorithm: Simulation with correct virtual input reconstruction, and experiment.}
    \label{fig:mi_check}
\end{figure}

The tracking performance of the mapping inversion algorithm is presented in Fig.~\ref{fig:mi_check}, both for the case of simulation and the experiment. The slight mismatch between the two can be related to model uncertainty, and to the fact that the digitally reproduced inertia is an equivalent load which wouldn't be present in the real turbine. Its effect on the total load $M_{h,l}$ is the addition of a high-frequency component which causes low-amplitude high-frequency oscillations in the signals, reflected in a virtual input displaying larger peaks. $v$ and $\rho$ are the signal defined in \eqref{eq:mi_cs}, while $v_{sim}$ is the equivalent of the $v$ signal which is obtained in simulation, hence based on the real states of the system and not on the noisy measurements. The algorithm is not able to perfectly reconstruct $v$, because of the presence of noise in the signals.

\begin{figure}[bp]
    \centering
\includegraphics[width=\columnwidth]{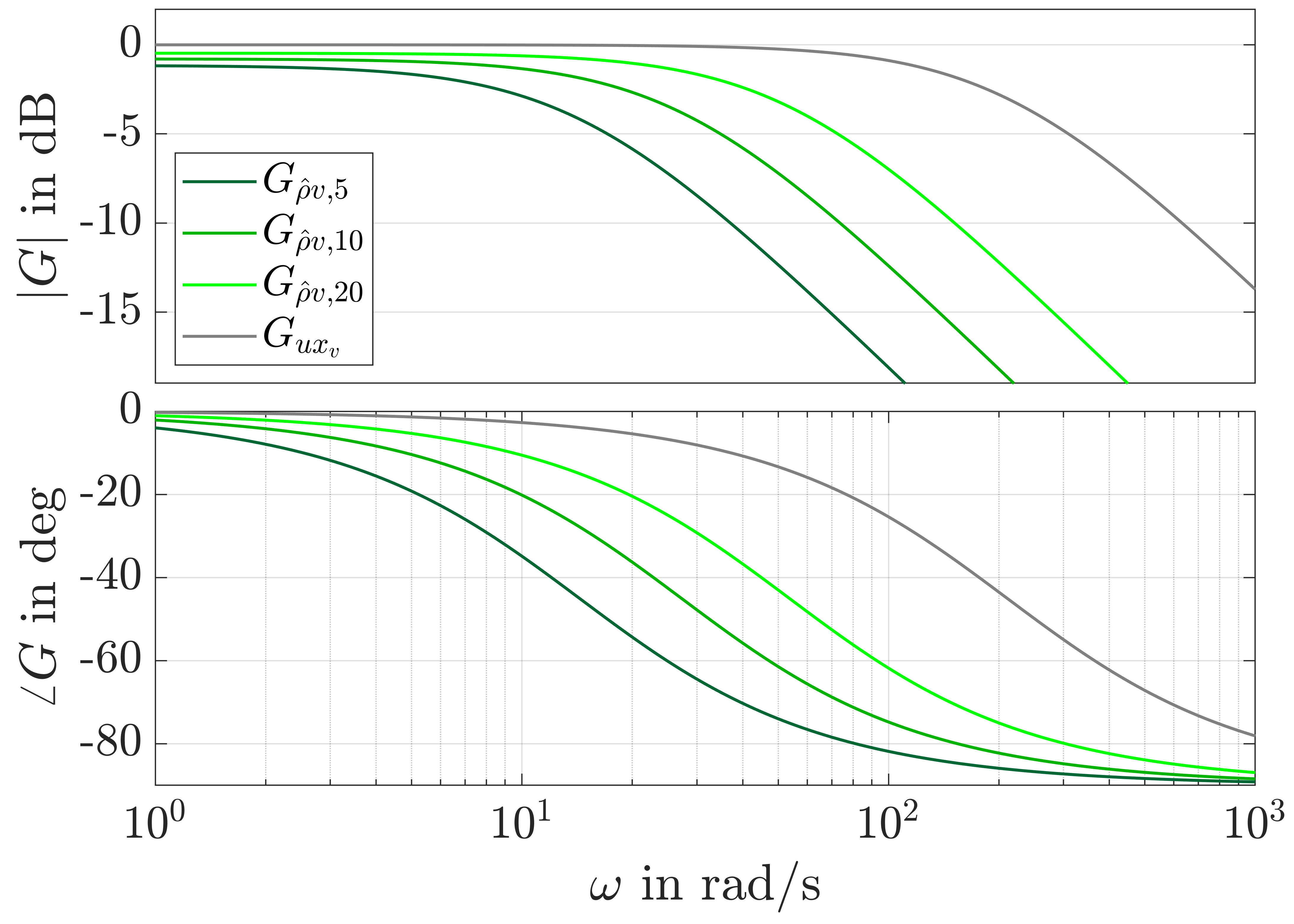}
    \caption{Bode plots of valve dynamics and mapping inversion filtering effect for different adaptation gains.}
    \label{fig:mi_bode}
\end{figure}

However, it is shown how it is possible to determine a proper tuning such that it is possible to reconstruct the actual virtual input in the plant ($v_{sim}$) and filtering out the noise present in the measurements. Thus, there is a natural filtering action which can be approximated with a simple first order low-pass filter in order to determine an indication of the bandwidth of such filtering action, in comparison to the bandwidth of the valve which represents a physical limitation in what the control action can deliver on the real system. The comparison, for $\gamma$ values of $(5,\,10,\,20)\times10^{-4}$, is presented in Fig.~\ref{fig:mi_bode}.

The cutoff frequency of the MI algorithm is always lower than the valve's bandwidth, meaning the valve only needs to follow trajectories it can actually withstand. With respect to the standard solution, where the filtering was occurring in the valve directly, in this new solution is the algorithm accounting for the valve limitations and filtering the input signal. This result support the similar signal quality in $u$ and $x_v$ in the MI solution that was found in Fig.~\ref{fig:fl_track_zoom}.

\begin{remark} \normalfont
    The tuning presented shows some room with respect to the maximum bandwidth that can be achieved by the valve. However, a faster tuning is reflected in a noisier reconstruction of $v_{sim}$ from the noisy measurements, leading to instability. The slew rate limitation of the valve may also play a role.
\end{remark}

\subsection{Controllers comparison}

While the plots are able to display some clear properties of the system and some qualitative performance indications can be obtained from them, a more rigorous analysis is performed to have a quantitative description and comparison of the proposed solutions. The controllers are tested according to some metrics, related both to the tracking performance and to the input cleanliness.

\begin{figure}[t]
    \centering
\includegraphics[width=1.03\columnwidth]{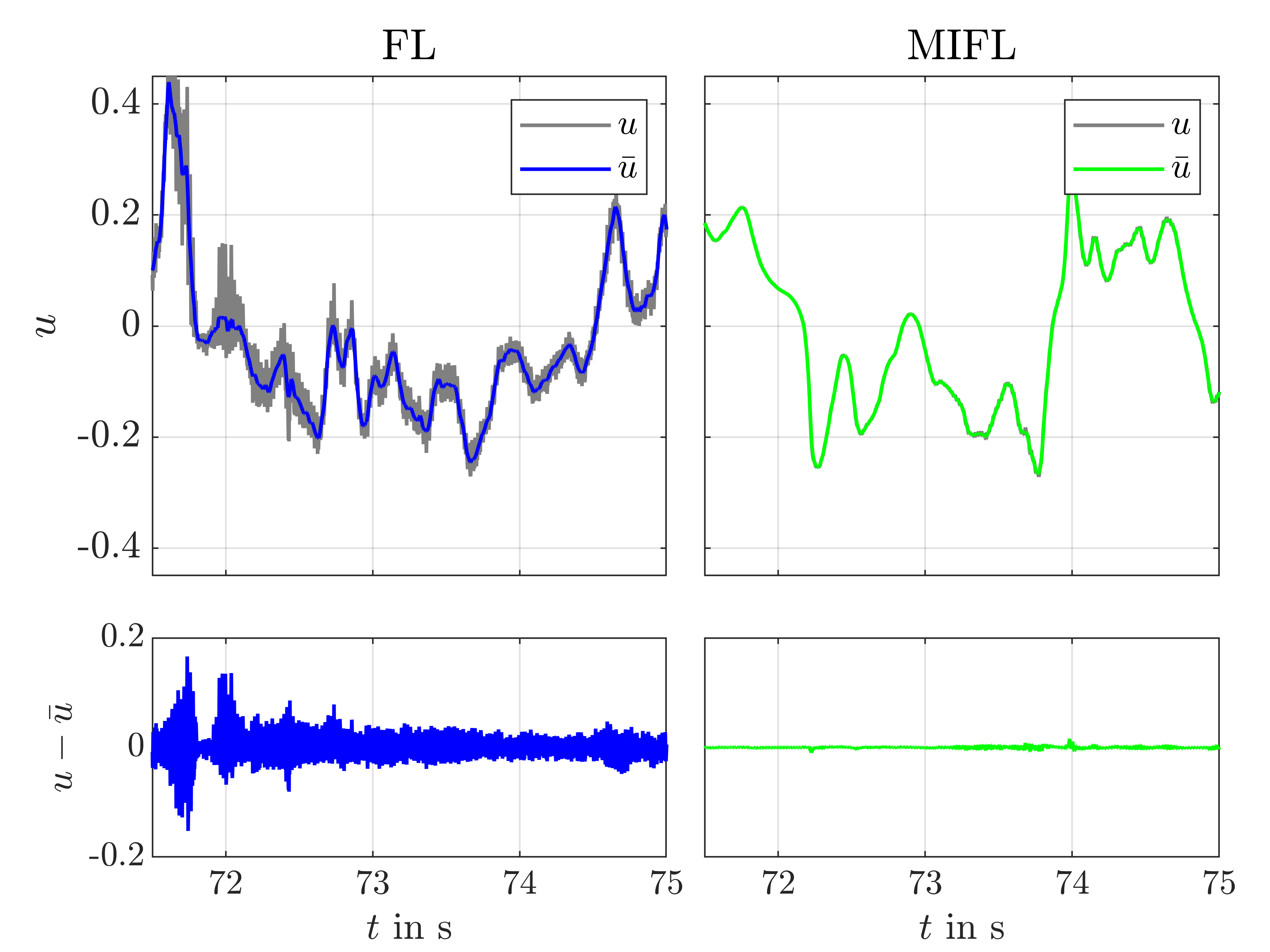}
    \caption{Input signal $u$ and moving average $\bar{u}$ (upper), and the difference $u-\bar{u}$ (lower). Left column plots are with FL alone, right column is with the MI combined with FL. Chattering is significantly reduced with inclusion of MI.}
    \label{fig:u_avg}
\end{figure}

Assessing the tracking performance based on the error is straightforward, while assessing the signal quality of the input is more challenging since it is not a constant average signal. The solution was to calculate a moving average $\Bar{u}$ and defining a new signal $u_f = u - \Bar{u}$ which has zero mean and still maintains the information of the high frequency oscillation. The moving average window is the same for all the tests. The same applies for determining $x_{v,f}$. The case of $u_f$, for the same moving average filter applied to both solutions, is illustrated in Fig.~\ref{fig:u_avg}.

The following metrics are defined for the purpose:
\setlength{\columnsep}{4pt}
\begin{multicols}{2}
\begin{itemize}
    \item \scalebox{0.85}{\vphantom{$\sqrt{\frac{\sum_{i=1}^{N} e_{\beta,i}^2 |e_{\beta,i}|}{\sum_{i=1}^{N} e_{\beta,i}^2}}$}%
    $\text{RMS}(u_f) = \sqrt{\frac{1}{N} \sum_{i=1}^{N} u_{fi}^2}$}

    \item \scalebox{0.85}{\vphantom{$\sqrt{\frac{\sum_{i=1}^{N} e_{\beta,i}^2 |e_{\beta,i}|}{\sum_{i=1}^{N} e_{\beta,i}^2}}$}%
    $\text{RMS}(x_{v,f}) = \sqrt{\frac{1}{N} \sum_{i=1}^{N} x_{v,fi}^2}$}

    \item \scalebox{0.85}{\vphantom{$\sqrt{\frac{\sum_{i=1}^{N} e_{\beta,i}^2 |e_{\beta,i}|}{\sum_{i=1}^{N} e_{\beta,i}^2}}$}%
    $\text{MAX}(e_{\beta}) = \max_{0\le t\le T} |e_{\beta}(t)|$}
\end{itemize}

\begin{itemize}
    \item \scalebox{0.85}{\vphantom{$\sqrt{\frac{\sum_{i=1}^{N} e_{\beta,i}^2 |e_{\beta,i}|}{\sum_{i=1}^{N} e_{\beta,i}^2}}$}%
    $\text{RMS}(e_{\beta}) = \sqrt{\frac{1}{N} \sum_{i=1}^{N} e_{\beta,i}^2}$}

    \item \scalebox{0.85}{$\text{QWAE}(e_{\beta}) = \sqrt{\frac{\sum_{i=1}^{N} e_{\beta,i}^2 |e_{\beta,i}|}{\sum_{i=1}^{N} e_{\beta,i}^2}}$}

    \item \scalebox{0.85}{\vphantom{$\sqrt{\frac{\sum_{i=1}^{N} e_{\beta,i}^2 |e_{\beta,i}|}{\sum_{i=1}^{N} e_{\beta,i}^2}}$}%
    $\text{ISE}(e_{\beta}) = \int_0^T e_{\beta}^2(t)\,dt$}
\end{itemize}
\end{multicols}
where $T$ is the simulation time and the resulting values are normalized for having a clearer comparison. The QWAE (Quadratically Weighted Average Error) metric is used to penalize larger errors more heavily, while the ISE (Integral Squared Error) metric measures the persistence of errors. Finally, the time window where all the controllers saturate is removed from the evaluation, since it is not representative of the real turbine operation.

\begin{figure}[bp]
    \centering
\includegraphics[width=\columnwidth]{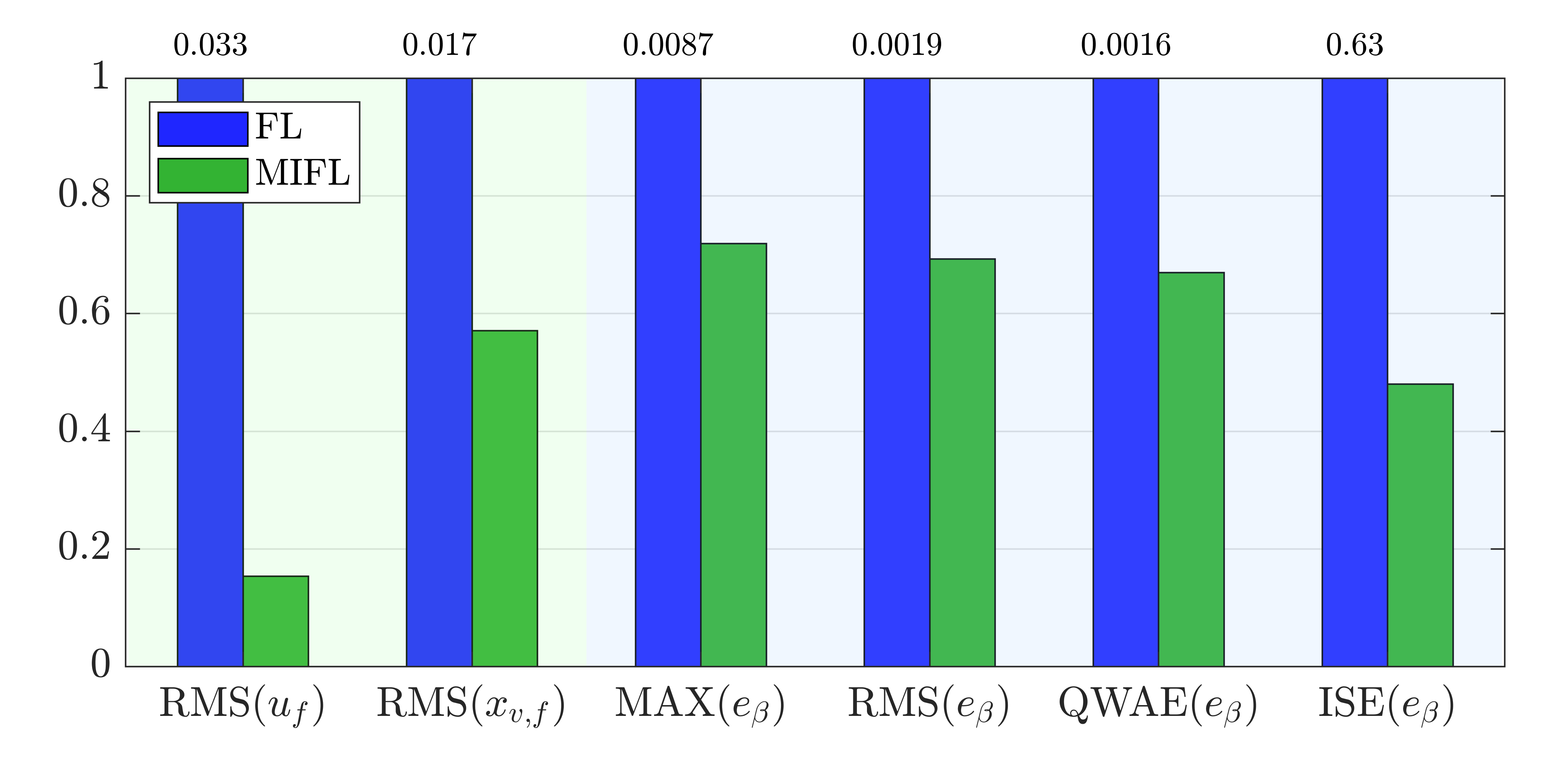}
    \caption{ Experimental results  comparing the baseline algorithm Fig. \ref{fig:block_diag_base} with the suggested MI approach Fig. \ref{fig:block_diag} . 
    Blue bars of the baseline control are normalized to 1, and green bars (scaled accordingly) show the quality measures for the MIFL solution, relative to that of the baseline approach (FL). Introduction of the the mapping inversion has very significant benefits. Normalization value presented on top of each group of bars.}
    \label{fig:metrics1}
\end{figure}

\begin{figure}[t]
    \centering
    \includegraphics[width=\columnwidth]{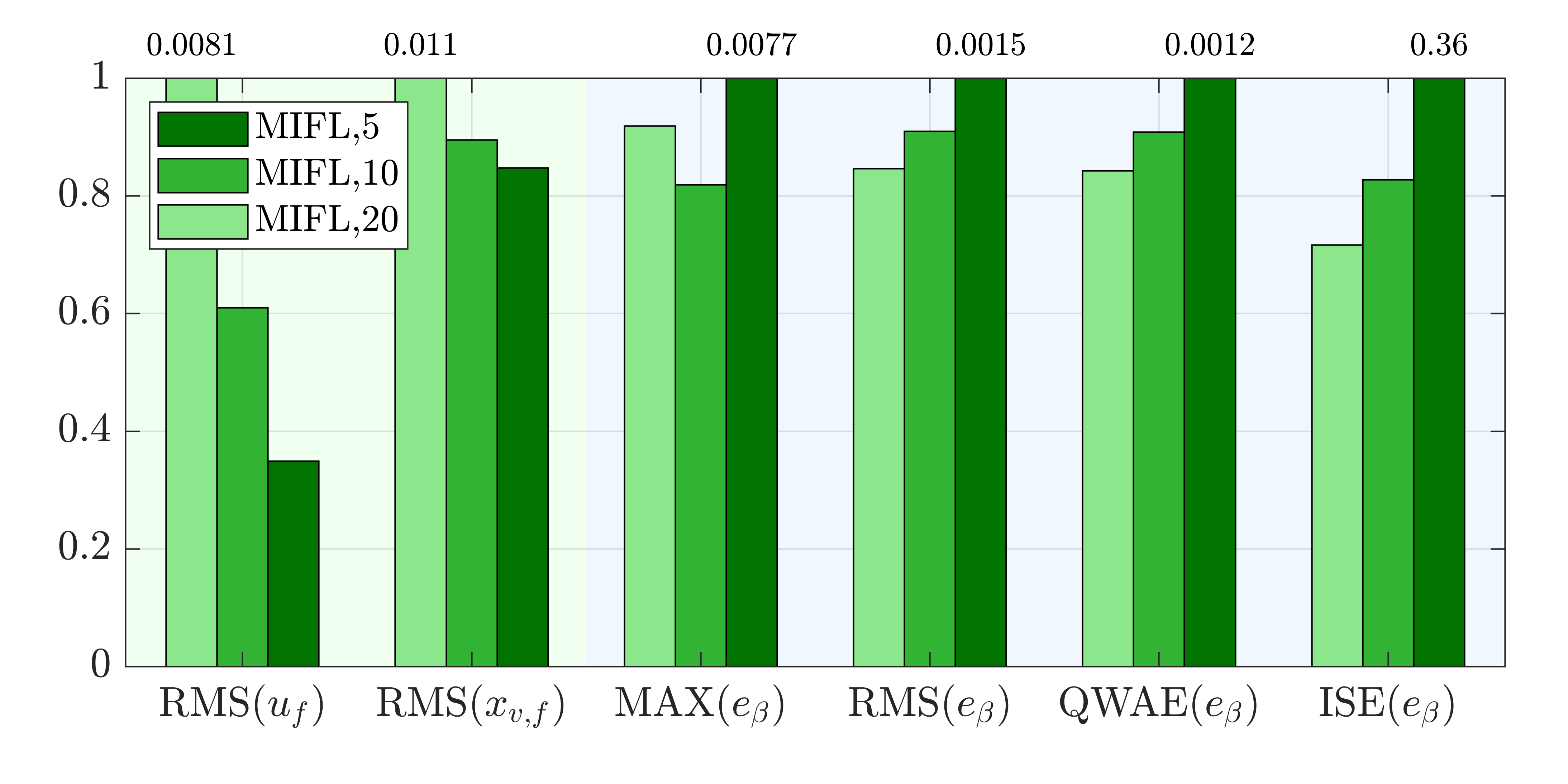}
    \caption{Experimental results comparing different tunings for the input inversion algorithm. Higher values of $\gamma$ are reflected in smaller error indexes, but larger chattering amplitude. Normalization values on top of each bar group is added to make comparison possible with Fig.\ref{fig:metrics1}.}
    \label{fig:metrics2}
\end{figure}

The experimental results for both controllers are presented in Fig.~\ref{fig:metrics1}, displaying how the solution proposed in the article outperforms the baseline for the same controller tuning and with the adaptation gain $\gamma = 10\times10^{-4}$, which corresponds to the tuning case $MIFL,10$ in Fig. \ref{fig:metrics2}. \hl{The adaptive law was implemented using forward Euler integration, with a sampling time of 1ms.}

The most evident effect is the chattering reduction in the actuator, which is lowered to less than 20$\%$ of the value for the baseline solution. The effect is less pronounced for the actual valve position, but the MI approach still reduces the energy in valve fluctuations down to less than 60$\%$, which can have a strong impact in the life cycle of the valve, since reduction of high-frequency cycling of the valve reduces the stress on the spool and wear induced by chattering.

On top of the improvement in the actuation, the Mapping Inversion solution shows an improvement in tracking performance, as seen in the metrics of the tracking error. A reduction of about 30$\%$ is obtained, with highest impact on ISE that shows a reduction of 50$\%$.

The results for three sets of tuning parameters of the inversion algorithm, see Fig.~\ref{fig:mi_bode}, are presented in Fig.~\ref{fig:metrics2}. Except for the maximum error metrics, all the others share the expected trend: error values decrease with higher values of $\gamma$, while input chatter values increase. The normalization value presented on top of each bar group is always smaller than the respective one presented in Fig.~\ref{fig:metrics1}, meaning that the solution with the MI outperforms the baseline FL for all the tuning cases of the MI.

The experimental results hence demonstrate that the MI approach, with the $\gamma$ tuning described, provides an easily implementable and effective solution that can obtain satisfactory tracking of an electro hydraulic actuator while obtaining significant reduction in wear caused by reduction in EHA micro movements and simultaneously by reduced energy in unwanted valve fluctuations.

\section{Conclusion}\label{sec:conclusions}

This article presented a solution to eliminate the formation of an algebraic loop when designing conventional nonlinear controllers. This was achieved by design of a virtual input and dynamically deriving a control action, which emulated the desired behavior, while control valve trajectory commands were avoided if they would exceed the specified admissible range for valve operations.

The effectiveness of the proposed solution was tested in high-fidelity simulation and experimentally for the industrial application of wind turbines' blade pitch control, showing a general improvement in  tracking performance and an essential reduction of input chatter, especially in the regions where the pitch reference is highly varying.

The dynamic inversion solution enabled room for flexibility and allows for future progress with other nonlinear parameter estimation algorithms. A further step towards industrial implementation will be to derive the \hl{implicit} discrete-time version of the algorithm.

\section*{Acknowledgements}
This research was funded by The Danish Energy Technology Development and Demonstration Programm (EUDP) through the project: “Decreased Cost of Energy (CoE) from wind turbines by reducing pitch system faults”, grant number 64022-1058. The authors much appreciate this support. The authors would also like to thank Prof. H. C. Pedersen, as well as postdoctoral researchers M. Bhola and G. Wrat, and the students at Aalborg University, for their valuable assistance in conducting the experiments.

\bibliographystyle{IEEEtran}
\bibliography{mi_bib}


 




\vfill

\end{document}